\documentclass[letterpaper,journal]{IEEEtran}

\usepackage{amsmath,amssymb,amsfonts}
\usepackage{array}
\usepackage{booktabs}
\usepackage{multirow}
\usepackage{graphicx}
\usepackage{siunitx}
\usepackage{cite}
\usepackage{url}
\usepackage{textcomp}
\usepackage{xcolor}
\usepackage{tikz}
\usetikzlibrary{positioning,arrows.meta,fit,backgrounds,decorations.pathreplacing,patterns}
\usepackage{pgfplots}
\pgfplotsset{compat=1.16}

\definecolor{dlcol}{RGB}{200,222,245}   
\definecolor{ulcol}{RGB}{250,222,196}   
\definecolor{scol}{RGB}{224,224,224}    
\definecolor{sndcol}{RGB}{ 66,133,205}  
\definecolor{symcol}{RGB}{198,226,208}  
\definecolor{ctrlcol}{RGB}{238,238,238} 

\DeclareSIUnit{\dBm}{dBm}
\DeclareSIUnit{\dBi}{dBi}

\newcolumntype{L}[1]{>{\raggedright\arraybackslash}p{#1}}
\newcommand{\code}[1]{\texttt{#1}}
\newcommand{\cmark}{$\checkmark$}
\newcommand{\xmark}{$\times$}

\definecolor{draftred}{RGB}{160,40,40}
\newif\ifdraftcontent
\draftcontentfalse

\tikzset{
  block/.style={draw,rounded corners,align=center,font=\footnotesize,
                minimum height=8mm,inner sep=3pt,fill=black!3},
  radar/.style={block,fill=blue!7,draw=blue!60!black},
  oran/.style={block,fill=green!8,draw=green!45!black},
  link/.style={-{Stealth[length=2mm]},thick},
  bilink/.style={{Stealth[length=2mm]}-{Stealth[length=2mm]},thick}
}

\begin{document}

\title{Real-Time Symbol-Domain OFDM Radar in an OpenAirInterface 5G Base Station With O-RAN Sensing Services}

\author{\IEEEauthorblockN{Karim Saifullin, \IEEEmembership{Graduate Student Member,~IEEE},  Sajid Ahmed, \IEEEmembership{Senior Member,~IEEE,} and Mohamed-Slim Alouini, \IEEEmembership{Fellow,~IEEE}}%
\thanks{Manuscript received Month XX, 2026; revised Month XX, 2026.}%
\thanks{\textit{Electrical and Computer Engineering (ECE)}\\{Computer, Electrical and Mathematical Sciences and Engineering (CEMSE)}
    \textit{King Abdullah University of Science and Technology (KAUST)},
    Thuwal, 23955-6900, Kingdom of Saudi Arabia\\
    karim.saifullin,sajid.ahmed,slim.alouini@kaust.edu.sa}%
}

\markboth{}%
{Author \MakeLowercase{\textit{et al.}}: Real-Time Symbol-Domain OFDM Radar in an OpenAirInterface 5G Base Station}

\maketitle

\begin{abstract}
This paper presents a real-time orthogonal frequency-division multiplexing
(OFDM) radar embedded in the OpenAirInterface (OAI) 5G base-station process.
The radar removes communication symbols by regularized element-wise division
and performs range--Doppler processing and ordered-statistic
constant-false-alarm-rate detection online without modifying the 5G waveform.
The implemented system provides \SI{2.57}{\meter} nominal range resolution
and \SI{0.28}{\meter\per\second} velocity resolution.
Hardware measurements identify and mitigate several implementation-specific
limitations, most notably a deterministic carrier-dependent transmit--receive
phase rotation on a Universal Software Radio Peripheral (USRP) X300. Selecting a tuning-grid-aligned carrier improves mean-removal
clutter suppression from \SI{-16.4}{\decibel} to \SI{38.0}{\decibel} and
reduces coherent-integration loss from \SI{19.81}{\decibel} to
\SI{0.27}{\decibel}. The measured processing gain closely agrees with its
predicted value. Instrumented worker timing confirms real-time operation,
with a conservative \SI{58.4}{\percent} utilization bound and no dropped
soundings. 
A custom E2 service model, E2SM-RADAR, exports detections and a compact
slow-time product to a near-real-time RAN Intelligent Controller. Live
end-to-end operation demonstrates reliable delivery and supports
controller-side tracking, micro-Doppler analysis, and classification. With a
commercial user equipment connected on the same carrier, 
measurements show no measurable difference in downlink throughput estimate
with sensing enabled, while the radar
sensing bandwidth follows the scheduler allocation.

\end{abstract}

\begin{IEEEkeywords}
5G, joint communication and sensing, integrated sensing and communication, OFDM radar, OpenAirInterface, O-RAN, radar systems, software-defined radio.
\end{IEEEkeywords}

\section{Introduction}
\IEEEPARstart{J}{oint} communication and radar (JCR) aims to use communication signals and infrastructure for sensing. A practical implementation must do more than show that a cellular waveform
can generate a range--Doppler map. The radar must operate within the base
station's timing and memory limits, use the scheduler-generated resource
grid, remain stable on radio hardware, and make its output available to
network applications. Many published testbeds address only part of this process. Some use standards-compliant waveforms but process recorded samples offline \cite{SW_invited_paper, oai_isac_ew25}. Others operate in real time using
communication-like orthogonal frequency-division multiplexing (OFDM) waveforms, but without an active cellular link
carrying user data through a complete protocol stack~\cite{adi_grofdmradar, fd_ofdm_radar_tmtt}. Recent Open Radio Access Network (O-RAN) studies expose channel estimates or spectral features, but not the output of a monostatic radar \cite{senseoran,batstation}.

This paper presents a complete implementation of symbol-domain OFDM radar inside the OpenAirInterface (OAI) 5G next-generation NodeB (gNB). The method follows the OFDM radar formulation introduced by Sturm, Wiesbeck, and co-authors~\cite{SW1,ofdm_sturm_first,SW_MLE,SW_invited_paper}. The received resource grid is divided by the transmitted grid, which removes the random payload symbols and leaves the target-dependent phase terms. The resulting matrix is processed across subcarriers for range and across sounded downlink slots for Doppler.

Prior work has demonstrated individual combinations of real-time OFDM radar,
payload-bearing sensing, OAI-based cellular operation, and O-RAN sensing
services. This work brings these capabilities together in a single
implementation: symbol-domain radar operates online inside an OAI 5G gNB on
real RF hardware, senses the payload-bearing downlink waveform, and exposes
the resulting sensing products through an O-RAN E2 service. The main
contributions are as follows:

\begin{itemize}
\item A real-time symbol-domain monostatic OFDM radar is implemented inside
the OAI 5G gNB. The radar operates directly on the downlink resource grid and
performs symbol removal, range compression, clutter removal, Doppler
processing, and constant-false-alarm-rate (CFAR) detection in real time on radio hardware. Operation is
demonstrated both with a controlled full-band configuration and with a
conventionally scheduled commercial user equipment (UE).

\item The deployed system is experimentally characterized, including a
carrier-dependent transmit--receive phase rotation observed on the Universal Software Radio Peripheral (USRP) X300
and its effect on clutter suppression and coherent integration. With the
corrected carrier configuration, measurements further evaluate processing
gain, false alarms, range consistency, and detector sensitivity.

\item The radar is exposed as an O-RAN sensing service through a custom E2
service model. Per-coherent processing interval (CPI) detections and a compact slow-time product are
transported to controller-side applications for monitoring, control,
tracking, micro-Doppler processing, recording, and classification.
\end{itemize}

The implementation is validated progressively in an offline physical-layer
simulator, the OAI radio frequency (RF) simulator, and the hardware gNB. This staged procedure
separates signal-processing errors from real-time software constraints and
radio-hardware effects.

\begin{table*}[bt]
\centering
\caption{Comparison With Closely Related OFDM Radar and Cellular Sensing Implementations}
\label{tab:comparison}
\footnotesize
\setlength{\tabcolsep}{4.2pt}
\begin{tabular}{L{3.05cm}cccccc}
\toprule
\textbf{Work} & \textbf{Symbol-domain radar} &
\textbf{OAI} & \textbf{Real targets\textsuperscript{a}} & \textbf{Online sensing} &
\textbf{Live payload sensing\textsuperscript{b}} & \textbf{O-RAN service}\\
\midrule
Sturm \emph{et al.}~\cite{SW_performance_verification}
& \cmark & \xmark  & \cmark & \xmark & \xmark & \xmark\\
ADI \code{gr-ofdmradar}~\cite{adi_grofdmradar}
& \cmark & \xmark  & \cmark & \cmark & \xmark & \xmark\\
Ozkaptan \emph{et al.}~\cite{mimo_jrc_mmwave}
& \cmark & \xmark & \cmark & \cmark & \cmark & \xmark\\
OpenISAC~\cite{openisac}
& \cmark & \xmark & \cmark & \cmark & \cmark & \xmark\\
CellSense~\cite{cellsense}
& \xmark & \cmark & \cmark & partial & \xmark & \xmark\\
Carbonara \emph{et al.}~\cite{oai_isac_ew25,oai_isac_jcs26}
& \cmark & \cmark & \xmark & \xmark & \cmark & \xmark\\
Bouknana \emph{et al.}~\cite{oran_loc_e2sm_srs}
& \xmark & \cmark & \xmark & \cmark & \xmark & \cmark\\
\textbf{This work}
& \cmark & \cmark & \cmark & \cmark & \cmark & \cmark\\
\midrule
\multicolumn{7}{L{17.1cm}}{\scriptsize
\textsuperscript{a} ``Real targets'' means physical objects that do not
  participate in the sensing process, observed over the air through their
  reflections.\newline
\textsuperscript{b} ``Live payload sensing'' means sensing from the
  payload-bearing downlink grid of an active communication link, rather
  than
  from a dedicated radar waveform, from reference signals alone, or from
  modulation symbols transmitted without a link partner.
}\\
\bottomrule
\end{tabular}
\end{table*}

The remainder of the paper is organized as follows.
Section~\ref{sec:related} reviews related OFDM radar and cellular sensing
systems. Section~\ref{sec:model} describes the signal model and processing
chain. Section~\ref{sec:implementation} presents the gNB implementation and
validation stages. Section~\ref{sec:oran} describes the O-RAN E2 interface
and controller-side applications. Section~\ref{sec:results} presents the
hardware, radar, and network-level measurements, and
Section~\ref{sec:conclusion} concludes the paper.

\section{Related Work and Positioning}
\label{sec:related}

Symbol-domain OFDM radar has been developed and validated through studies on range--Doppler
processing, maximum-likelihood estimation, and hardware implementation
\cite{SW1,ofdm_sturm_first,SW_MLE,SW_performance_verification,SW_invited_paper}.
Subsequent research has extended this framework to address interference cancellation, multiuser operation, multiple-input multiple-output (MIMO) processing, sparse reference-signal placement, and full-duplex transceiver architectures
\cite{IC_idea_first,ofdm_multiuser,ofdm_mimo_angular,PRS_fundamentals,
fd_ofdm_radar_tmtt}.

Several software-defined radio (SDR) platforms have demonstrated real-time OFDM radar outside a
3GPP base-station stack. Examples include the GNU Radio
\code{gr-ofdmradar} implementation~\cite{adi_grofdmradar}, a real-time
mmWave joint radar--communication platform~\cite{mimo_jrc_mmwave}, and
OpenISAC~\cite{openisac}. The latter two systems also demonstrate sensing
using communication-bearing waveforms, but do not integrate the radar
processing into a live 3GPP gNB stack.

Other studies have instead built sensing functions directly on OAI.
CellSense uses uplink sounding reference signals for passive angle--delay
sensing~\cite{cellsense}, while Carbonara \emph{et al.} investigated
self-interference mitigation and symbol-domain OFDM radar using 5G-compliant
OAI signals on real radio hardware, with radar processing performed on
recorded data or using a channel emulator~\cite{oai_isac_ew25,oai_isac_jcs26}. The present work combines
these previously separate capabilities by performing symbol removal,
range--Doppler processing, moving-target indication (MTI), and CFAR detection
continuously inside the running OAI gNB while sensing the payload-bearing
downlink waveform. The implementation further addresses real-time integration
constraints, including bounded buffering and isolation from slot-critical
processing, together with radio-hardware coherence.

Related work has also considered sensing functions within O-RAN. Some studies
place sensing functions close to an OAI Distributed Unit through
dApps~\cite{dapp_framework,dapp_isac}, while others introduce
sensing-specific interfaces. E2SM-SENS exports spectral sensing features,
whereas SenseORAN and BatStation address incumbent-radar detection and
spectrum control rather than target range--Doppler sensing
\cite{oran_isac_arch_analysis2,senseoran,batstation}. Bouknana \emph{et al.}
use a custom E2 service model on the OAI--FlexRIC stack to export uplink
sounding reference signal (SRS) channel estimates to the near-RT RIC, where
an xApp performs continuous inference using a channel-charting model to
localize connected user equipment~\cite{oran_loc_e2sm_srs}. Their framework
and the present one share the use of a custom E2 service model for real-time
sensing information, but differ in sensing modality: their system estimates
the position of a cooperative UE from its own uplink transmissions, whereas
the present system reports monostatic range--Doppler detections of
uncooperative targets obtained from the downlink resource grid.

Table~\ref{tab:comparison} summarizes the capabilities of the most closely
related implementations and positions the present system with respect to
online OFDM radar, OAI integration, payload-bearing sensing, and O-RAN
service exposure. The signal model and processing chain are described next.

\section{Symbol-Domain OFDM Radar}
\label{sec:model}

\subsection{Signal Model}
Consider $M$ sounded downlink slots separated by the slow-time interval $T_{\mathrm{s}}$. Each sounded slot contains $L_{\mathrm{s}}$ OFDM symbols and $N$ used subcarriers. Let $X[m,\ell,n]$ denote the transmitted modulation symbol in sounded slot $m$, OFDM symbol $\ell$, and subcarrier $n$. For $K$ point targets, the corresponding received frequency-domain sample can be written as
\begin{equation}
\begin{aligned}
Y[m,\ell,n] ={}& X[m,\ell,n]\sum_{i=0}^{K-1}\alpha_i\\
&{}\times e^{-j2\pi n\Delta f\tau_i}
 e^{j2\pi f_{D,i}(mT_{\mathrm{s}}+t_\ell)} + W[m,\ell,n],
\end{aligned}
\label{eq:rxmodel}
\end{equation}
where $\alpha_i$ is the complex target coefficient, $\tau_i=2R_i/c_0$ is the round-trip delay, $f_{D,i}=2v_i f_c/c_0$ is the Doppler frequency, $\Delta f$ is the subcarrier spacing, $t_\ell$ is the symbol time within the sounded slot, and $W[m,\ell,n]$ contains receiver noise and residual interference. The cyclic prefix must be longer than the target delay for the model to hold.

The implementation removes the transmitted symbols by regularized conjugate division. Let
\begin{equation}
\overline P_X[m]=\frac{1}{|\mathcal{O}_m|}
\sum_{(\ell,n)\in\mathcal{O}_m}|X[m,\ell,n]|^2
\end{equation}
be the mean transmitted resource-element power in sounded slot $m$, where
$\mathcal{O}_m$ contains the occupied resource elements of that slot,
excluding the guard bands, the DC subcarrier and its associated guard, and
all elements that are exactly zero in the transmitted grid.

To protect the system from numerical instability, we define the 
relative validity mask, which excludes low-energy elements, and the
corresponding per-resource-element estimate as
\begin{align}
\chi[m,\ell,n] &={}
\mathbb{I}\!\left\{|X[m,\ell,n]|^2\geq
\eta\overline P_X[m]\right\},
\label{eq:remask}\\
Z[m,\ell,n] &={}
\chi[m,\ell,n]
\frac{Y[m,\ell,n]X^{*}[m,\ell,n]}
{|X[m,\ell,n]|^2+\epsilon}.
\label{eq:division}
\end{align}
Here $\epsilon$ regularizes the denominator and the relative mask uses
$\eta=0.1$ in the reported experiments. Guard-band subcarriers are also excluded.

\begin{figure*}[bt] \centering \resizebox{\textwidth}{!}{ \begin{tikzpicture}[node distance=7mm and 10mm] \node[block] (scene) {Target and clutter}; \node[block,right=of scene] (usrp) {USRP X300\\separate transmit (TX)/receive (RX) daughterboards}; \node[block,right=of usrp] (phy) {OAI gNB PHY\\scheduler and OFDM chain}; \node[radar,below=of phy] (ring) {Bounded snapshot ring\\TX/RX slot data}; \node[radar,right=of ring] (worker) {Radar worker\\division, range IFFT, MTI,\\Doppler FFT, OS-CFAR}; \node[oran,right=of worker] (agent) {OAI E2 agent\\E2SM-RADAR}; \node[oran,right=of agent] (ric) {Near-RT RIC\\display, record, control,\\tracking, classification}; \draw[bilink] (scene) -- node[above,font=\footnotesize]{echo} (usrp); \draw[bilink] (usrp) -- node[above,font=\footnotesize]{IQ} (phy); \draw[link] (phy) -- node[right,font=\footnotesize]{copy only} (ring); \draw[link] (ring) -- (worker); \draw[link] (worker) -- (agent); \draw[bilink] (agent) -- (ric); \end{tikzpicture} } \caption{Implemented architecture. The gNB real-time thread only copies slot data into a ring. All radar processing runs in a background worker. The E2 agent reads a per-CPI report and does not access the real-time radio state.} \label{fig:system_arch} \end{figure*}
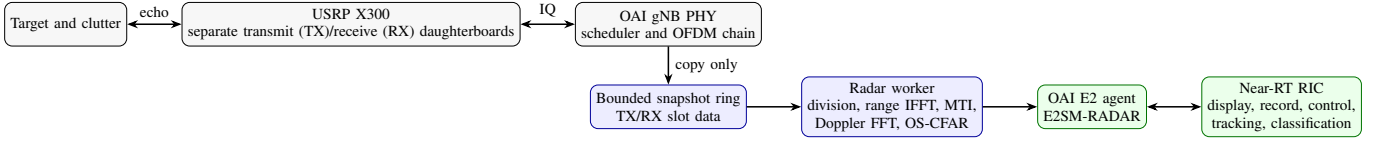

Range processing is performed separately for every sounded slot. The
$N$ occupied channel estimates are mapped to their corresponding bins of an
$N_{\mathrm{FFT}}$-point frequency grid, with the unused bins set to zero.
Let $\overline Z_F[m,k]$ denote this zero-padded grid. The range profile is
then
\begin{equation}
h[m,p]=
\sum_{k=0}^{N_{\mathrm{FFT}}-1}
w_R[k]\overline Z_F[m,k]
e^{j2\pi kp/N_{\mathrm{FFT}}},
\label{eq:rangeprofile}
\end{equation}
where $w_R[k]$ is the range window and $N_{\mathrm{FFT}}=2048$. Let $a[m]\in\{0,1\}$ denote the validity of slow-time sample $m$. Static coupling and clutter are removed at each range bin using
\begin{equation}
\widetilde h[m,p]=a[m]\left(h[m,p]-
\frac{\sum_{r=0}^{M-1}a[r]h[r,p]}{\sum_{r=0}^{M-1}a[r]}\right).
\label{eq:mti}
\end{equation}
Invalid soundings are excluded from the clutter estimate and set to zero
after mean removal, as represented by $a[m]$ in~\eqref{eq:mti}. A Doppler window and discrete Fourier transform across slow time then produce
\begin{equation}
P[p,q]=\left|\sum_{m=0}^{M-1}w_D[m]\widetilde h[m,p]
e^{-j2\pi mq/M}\right|^2.
\label{eq:rdmap}
\end{equation}
A two-dimensional ordered-statistic (OS-CFAR) detector is applied to $P[p,q]$. The implementation reports the range bin, Doppler bin, physical range, radial velocity, power, and estimated signal-to-noise ratio for each detection.

\subsection{Resolution and Ambiguity}
The basic range and velocity limits are
\begin{align}
\Delta R &= \frac{c_0}{2B},
& R_{\mathrm{amb}} &= \frac{c_0}{2\Delta f},
\label{eq:range_metrics}\\
\Delta v &= \frac{c_0}{2f_cMT_{\mathrm{s}}},
& v_{\mathrm{amb}} &= \frac{c_0}{4f_cT_{\mathrm{s}}}.
\label{eq:velocity_metrics}
\end{align}
Here $B=N\Delta f$ is the occupied sensing bandwidth. The nominal range
resolution is $c_0/(2B)$, whereas the 2048-point inverse FFT samples the
delay axis at $c_0/(2f_s)$. For the measurement configuration in
Table~\ref{tab:operating_point}, these values are
\SI{2.57}{\meter} and \SI{2.44}{\meter}, respectively, so the range
profile is oversampled by a factor of approximately 1.05.

With conventional scheduling, the available sensing bandwidth is determined
by the occupied downlink resource elements and can therefore vary with the
scheduler allocation. The corresponding range resolution degrades when the
occupied bandwidth is reduced. The maximum useful range is additionally
limited by the cyclic prefix, the receiver processing convention, and the
link budget.

\section{System Design and Implementation}
\label{sec:implementation}

\subsection{Platform Selection}

The hardware platform uses a single USRP X300 with transmit and receive on separate daughterboards. An initial two-radio configuration sharing an external frequency and timing reference did not improve leakage isolation when the antennas were connected, provided no short-term coherence advantage, and introduced an additional inter-device synchronization path. The single-radio arrangement was therefore selected for the final implementation. The N310 provides an independent reference and exhibits negligible rotation
at the tested carriers.

\subsection{Architecture}
Figure~\ref{fig:system_arch} shows the implemented system. The radar module is
self-contained and compiled into the OAI PHY library~\cite{OAI_foundational}. OAI generates the
downlink resource grid and time-domain transmit samples, while the real-time
radio thread only copies the required transmit and receive data into a bounded
ring. A background worker reads the ring, reconstructs the aligned grids,
executes the processing chain described in Section~\ref{sec:model}, and publishes one
report per CPI. The E2 agent forwards this report to the near-real-time RAN Intelligent Controller (RIC),
while runtime controls are returned through a mailbox and applied at CPI
boundaries. The same processing module is used in simulation and hardware,
with only the source of the transmit and receive grids changing between
validation stages.

\begin{table}[ht]
\centering
\caption{JCR Measurement Configuration}
\label{tab:operating_point}
\footnotesize
\begin{tabular}{ll}
\toprule
Parameter & Value\\
\midrule
Carrier frequency & \SI{3.34848}{\giga\hertz}\\
Subcarrier spacing & \SI{30}{\kilo\hertz}\\
Allocated resource blocks & 162\\
Occupied sensing bandwidth & \SI{58.29}{\mega\hertz}\\
Nominal range resolution & \SI{2.57}{\meter}\\
Range-bin spacing & \SI{2.44}{\meter}\\
Implementation ISI-free span & \SI{307}{\meter} (126 bins)\\
Nominal CP-limited span & approximately \SI{351}{\meter}\\
Slow-time interval & \SI{2.5}{\milli\second}\\
Slow-time samples per CPI & 64\\
CPI duration & \SI{160}{\milli\second}\\
Velocity resolution & \SI{0.28}{\meter\per\second}\\
Unambiguous velocity & $\pm\SI{8.95}{\meter\per\second}$\\
Detector & 2-D ordered-statistic CFAR\\
\bottomrule
\end{tabular}
\end{table}

\subsection{Real-Time Data Path, Grid Alignment, and CPI Formation}
\label{sec:datapath}

The gNB provides both the scheduled transmit resource grid and the
corresponding time-domain radio samples. Radar processing reconstructs the
transmit and receive grids using matched cyclic-prefix removal, Fourier
transform, subcarrier mapping, and phase conventions, since any mismatch can
introduce deterministic phase errors that appear as spurious range or Doppler
shifts.

For each downlink slot, the implementation stores a compact snapshot together with its frame, slot, time-division duplexing (TDD), and sounding-grid indices. A CPI is processed only when at least 50\% of the expected soundings are valid; thus, for $M=64$, at least 32 soundings are required. Missing or corrupted receive slots are marked invalid and excluded from the
clutter estimate rather than treated as valid zero-valued measurements. After
mean removal, their slow-time samples are set to zero through the validity
mask in~\eqref{eq:mti}. Hardware measurements also revealed occasional
receive dropouts, which are rejected using an occupied-band energy threshold
of 0.05 relative to an exponential moving average over accepted soundings.

The real-time thread performs only bounded copies and counter updates, while all radar processing is delegated to a worker thread. Completed CPIs are published as immutable report snapshots. This separation keeps radar processing outside the gNB scheduling deadline and exposes overload through validity and ring-drop counters. No ring drops were observed in the reported hardware experiments.

\begin{figure}[ht]
    \centering
    \begin{tikzpicture}[font=\sffamily\scriptsize,>={Latex[length=4pt]}]

\def\sw{0.42}          
\def\sh{0.62}          

\node[anchor=west] at (0,\sh+0.42) {\textbf{(a) One 10\,ms frame $=$ 20 slots \; (band n78, $\mu=1$, 30\,kHz SCS, 0.5\,ms/slot)}};

\foreach \s in {0,1,2,3,4,5,6,10,11,12,13,14,15,16}
  \filldraw[fill=dlcol,draw=black!55] (\s*\sw,0) rectangle ++(\sw,\sh);
\foreach \s in {7,17}
  \filldraw[fill=scol,draw=black!55] (\s*\sw,0) rectangle ++(\sw,\sh);
\foreach \s in {8,9,18,19}
  \filldraw[fill=ulcol,draw=black!55] (\s*\sw,0) rectangle ++(\sw,\sh);

\foreach \s in {0,1,2,3,4,5,6,10,11,12,13,14,15,16}
  \node at (\s*\sw+0.5*\sw,\sh-0.20) {\tiny D};
\foreach \s in {7,17} \node at (\s*\sw+0.5*\sw,\sh-0.20) {\tiny S};
\foreach \s in {8,9,18,19} \node at (\s*\sw+0.5*\sw,\sh-0.20) {\tiny U};

\foreach \s in {1,6,11,16}{
  \filldraw[fill=sndcol,draw=black,line width=0.5pt] (\s*\sw,0) rectangle ++(\sw,\sh);
  \node[white] at (\s*\sw+0.5*\sw,\sh-0.22) {\tiny\bfseries D};
  \draw[->,thick,sndcol!85!black] (\s*\sw+0.5*\sw,\sh+0.30) -- (\s*\sw+0.5*\sw,\sh+0.04);
}

\foreach \s in {0,2,4,6,8,10,12,14,16,18}
  \node at (\s*\sw+0.5*\sw,-0.20) {\tiny \s};
\node[anchor=west] at (5.5,-0.52) {\tiny slot index within the frame};

\draw[->,black!60] (0.9,-0.6) -- (0.5*\sw,-0.05);
\node[anchor=west,black!60] at (0.2,-0.70) {\tiny slot 0 carries the SSB $\Rightarrow$ skipped (sound\_offset $=1$)};

\draw[decorate,decoration={brace,amplitude=3pt},black!70]
      (6*\sw+\sw,\sh+0.72) -- (1*\sw,\sh+0.72)
      node[midway,above=1pt,black!70] {\tiny PRI $= 5$ slots $= 2.5$\,ms};

\begin{scope}[shift={(0,-2.85)}]
\def\yw{0.60}          
\def\yh{0.70}          

\node[anchor=west] at (0,\yh+0.40) {\textbf{(b) Inside one sounded slot: all 14 OFDM symbols are used}};

\foreach \l in {1,...,13}
  \filldraw[fill=symcol,draw=black!55] (\l*\yw,0) rectangle ++(\yw,\yh);
\filldraw[fill=ctrlcol,draw=black!55] (0,0) rectangle ++(\yw,\yh);

\foreach \l in {0,...,13}
  \node at (\l*\yw+0.5*\yw,\yh-0.22) {\tiny $\ell{=}\l$};
\draw[|-|,black!60] (0,-0.16) -- (0.09,-0.16);
\node[anchor=west,black!60] at (0.10,-0.16) {\tiny CP};

\node[anchor=west,align=left] at (0,-0.55)
  {\tiny Valid resource elements from all 14 OFDM symbols contribute to the sounded slot.};


\draw[dashed,black!55] (1*0.42,2.85) -- (0,\yh);
\draw[dashed,black!55] (2*0.42,2.85) -- (14*\yw,\yh);
\end{scope}

\begin{scope}[shift={(0,-5)}]
\def\cw{0.13125}       
\def\ch{0.55}

\node[anchor=west] at (0,\ch+0.40) {\textbf{(c) One CPI: $M=64$ sounded slots over 160\,ms}};

\foreach \f in {0,...,16}
  \draw[black!25,dashed,line width=0.3pt] (\f*4*\cw,-0.10) -- (\f*4*\cw,\ch+0.14);

\foreach \m in {0,...,63}
  \filldraw[fill=sndcol,draw=sndcol!60!black,line width=0.2pt]
    (\m*\cw+0.028,0) rectangle ++(\cw-0.056,\ch);


\node[anchor=west] at (0,-0.3) {\tiny $m=0$};
\node[anchor=east] at (8.4,-0.3) {\tiny $m=63$};

\end{scope}

\begin{scope}[shift={(0,-1.30)}]
  \filldraw[fill=dlcol,draw=black!55] (0,0) rectangle ++(0.34,0.30);
  \node[anchor=west] at (0.40,0.15) {\tiny DL slot};
  \filldraw[fill=scol,draw=black!55] (1.45,0) rectangle ++(0.34,0.30);
  \node[anchor=west] at (1.85,0.15) {\tiny S slot (6D\,+\,4G\,+\,4U sym)};
  \filldraw[fill=ulcol,draw=black!55] (4.60,0) rectangle ++(0.34,0.30);
  \node[anchor=west] at (5.00,0.15) {\tiny UL slot};
  \filldraw[fill=sndcol,draw=black] (6.10,0) rectangle ++(0.34,0.30);
  \node[anchor=west] at (6.50,0.15) {\tiny sounded (radar) slot};
\end{scope}

\end{tikzpicture}
   \caption{Temporal mapping from the TDD frame to one radar CPI. Four selected downlink slots per \SI{10}{\milli\second} frame provide a \SI{2.5}{\milli\second} slow-time interval. Each selected slot produces one range profile, and 64 profiles form a \SI{160}{\milli\second} CPI for Doppler processing and CFAR detection. The radar reuses existing downlink transmissions and introduces no dedicated sensing air-interface resources.}
    \label{fig:radar_time_resources}
\end{figure}
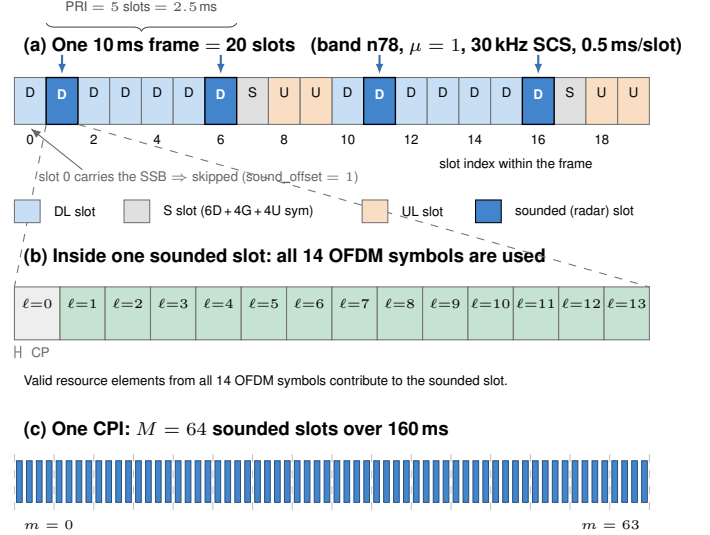

The sensing schedule maps selected downlink slots to slow-time samples for Doppler processing, as illustrated in Figure~\ref{fig:radar_time_resources}. 

The applied TDD configuration contains 20 slots per \SI{10}{\milli\second} frame, following a $7\mathrm{D}+1\mathrm{S}+2\mathrm{U}$ pattern repeated every \SI{5}{\milli\second}. The radar selects downlink slots satisfying
\begin{equation}
s_{\mathrm{abs}} \bmod 5 = 1,
\end{equation}
which provides four soundings per frame at a \SI{2.5}{\milli\second} slow-time interval and avoids slot~0, which may contain the synchronization signal block.

Each selected slot produces one range profile. A CPI contains $M=64$ profiles
and spans \SI{160}{\milli\second}; the resulting profiles are processed using
the clutter-removal, Doppler, and detection chain in
Section~\ref{sec:model}. The radar therefore reuses existing downlink
transmissions without utilizing dedicated sensing air-interface resources.

\subsection{Numerical Processing and Detection}

Strong direct transmit--receive coupling creates a large dynamic-range
requirement, making numerical conditioning important. In a noiseless
two-target simulation, fixed-point processing limited the peak-to-median
dynamic range to \SI{37.5}{\decibel}, compared with
\SI{84}{\decibel} for the corresponding floating-point chain. The aligned
transmit and receive grids are therefore converted to single-precision
complex values before symbol division, and all subsequent radar processing is
performed in floating point.

To protect the symbol-division stage from numerical instability and noise
enhancement caused by small transmit amplitudes, we combine denominator
regularization with relative masking. The regularizer $\epsilon$ prevents
division by zero, while the mask in~\eqref{eq:remask} removes small nonzero
transmit samples relative to the power scale of each slot. In hardware
captures from both radios, disabling the mask raised the post-MTI Doppler
noise floor by \SIrange{17}{33}{\decibel}.

To obtain robust target detections while limiting responses to nearby clutter
and sidelobes, the two-dimensional OS-CFAR detector combines local
ordered-statistic thresholding with peak selection and clustering. It uses two
guard cells and eight training cells on each side along both range and Doppler,
with Doppler training cells wrapped at the velocity boundary. Candidate
detections must be local maxima within a $3\times3$ neighborhood and are
clustered within eight range bins and four Doppler bins. The reported
signal-to-noise ratio (SNR) is referenced to the local ordered-statistic
background estimate and is therefore a detector-relative quantity rather than
a calibrated echo SNR.

\subsection{Prototype, Simulation, and Validation Stages} 

The radar was developed through a sequence of validation stages. We first
adapted and substantially extended a GNU Radio prototype~\cite{gnuradio},
which did not originally operate on USRP hardware, by modifying both the
radio-interface code and the signal-processing chain. The resulting prototype
was used to verify the division-based estimator, basic range--Doppler
processing, radio topology, and the effect of payload modulation.

The processing module was then implemented in an offline OAI physical-layer simulator, where targets with known delay and Doppler were inserted into the transmitted waveform. This stage verified the OAI resource-grid mapping and the transmit--receive phase conventions required before symbol-domain division. 

Finally, the same module was integrated into \code{nr-softmodem}, exercised with the OAI RF simulator, and deployed on radio hardware. Intermediate range and Doppler outputs were compared with an independent floating-point Python implementation at each stage. Recorded hardware slots could also be replayed through the processing chain, allowing algorithmic and parameter changes to be evaluated without repeating the radio measurement.

\subsection{Hardware Integration}
\begin{figure}[!t]
  \centering
  \includegraphics[width=0.9\columnwidth]{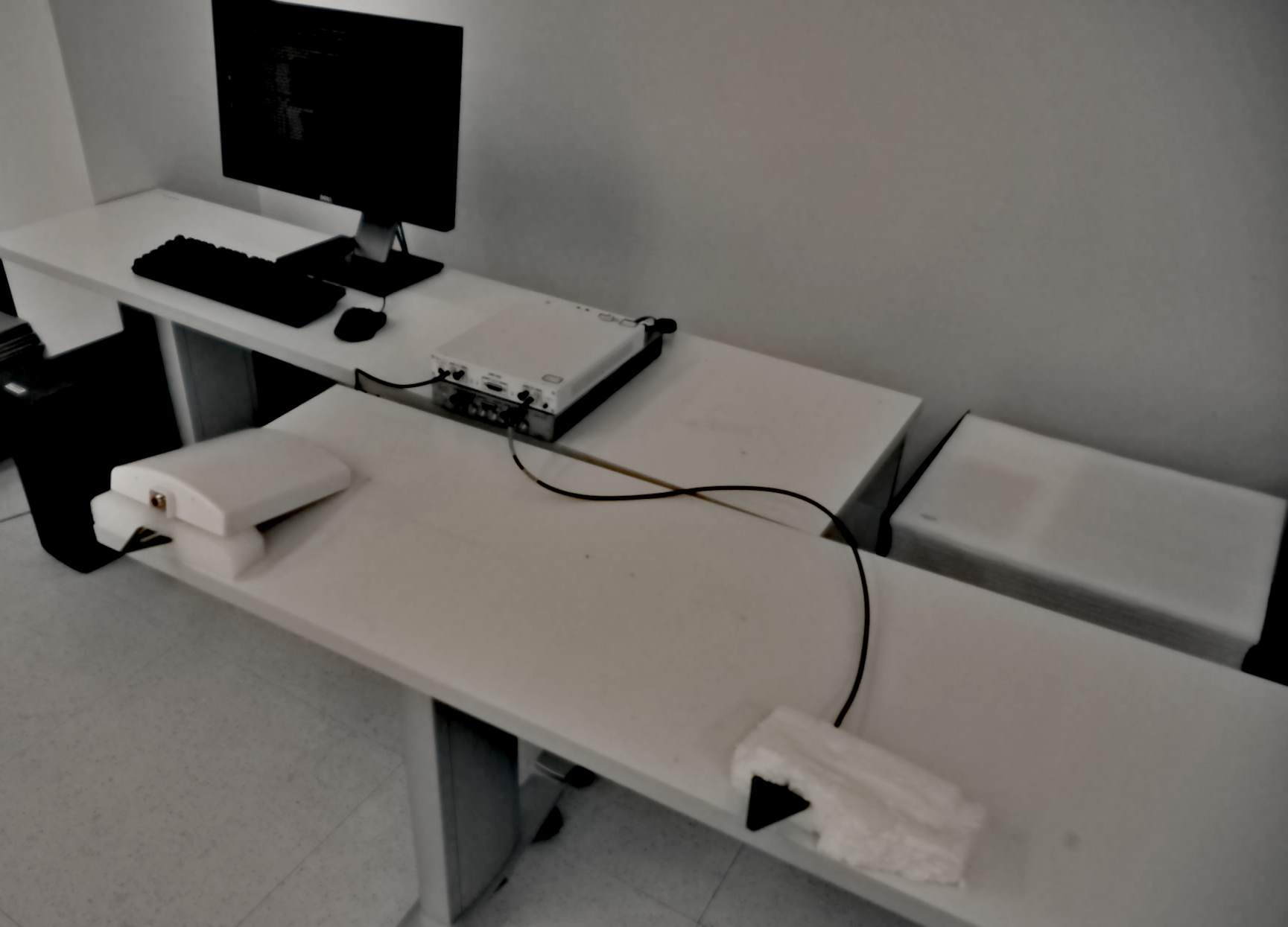}
  \caption{Hardware setup for the monostatic OAI radar experiments. The
  USRP X300 uses separate transmit and receive paths, with the antennas
  oriented along the measured target path.}
  \label{fig:hardware_setup}
\end{figure}

The hardware implementation uses a USRP X300 with transmit and receive on
separate daughterboards. Figure~\ref{fig:hardware_setup} shows the hardware
setup and antenna arrangement. The radar can operate with the conventional OAI
scheduler, in which the available sensing bandwidth follows the scheduled
downlink allocation. In this case, partial allocation can reduce the sensing
bandwidth and degrade radar performance. Therefore, for controlled hardware
characterization, the main radar measurements use OAI
\code{--phy-test}, which provides a full-band downlink allocation with no UE
attached. The corresponding transmit and receive samples are copied from the
real-time path through the bounded, nonblocking ring described above.

Hardware deployment exposed two system-level effects not present in
simulation. First, the \SI{500}{\micro\second} slot deadline required radar
processing to remain outside the real-time radio thread. Second, a
deterministic transmit--receive phase rotation degraded clutter cancellation
and coherent integration. The timing constraint is addressed by the
background-worker architecture, while the coherence impairment is examined in
Subsection~\ref{sec:coherence}.

To reduce direct transmit--receive coupling, the TX and RX antennas were
separated by approximately \SI{1}{\meter}, resulting in a quasi-monostatic
sensing geometry. The antenna arrangement was selected based on both target
return and direct coupling rather than isolation alone. Direct air coupling
was approximately \SI{-39}{\decibel}. Cross-polarization improved isolation
by up to \SI{28}{\decibel}, but co-polarized antennas produced the strongest
target-to-noise response for the tested short-range reflector geometry.
The final configuration therefore balances coupling headroom against target
illumination. No analog self-interference canceller is used.

\subsection{Transmit--Receive Coherence}
\label{sec:coherence}

\subsubsection{Observed Phase Rotation}

The direct transmit--receive coupling is a static component and should
therefore remain at zero Doppler. However, at a carrier of
\SI{3.32976}{\giga\hertz}, its phase advanced at approximately
\SI{28.57}{\hertz}, corresponding to an apparent radial velocity of
\SI{1.286}{\meter\per\second}. The rotation displaced the static coupling
from zero Doppler, degraded mean-based clutter removal, and reduced coherent
integration.

The effect was repeatable across two X300 devices at a common carrier.
The rotation rate varied with operating frequency in both magnitude and sign.
For example, \SI{3.60000}{\giga\hertz} and
\SI{3.61920}{\giga\hertz} produced rotations of
\SI{+6.5914}{\hertz} and \SI{-6.594}{\hertz}, respectively. This
device-repeatable but carrier-dependent behavior indicates a deterministic
tuning effect rather than independent oscillator instability.

\subsubsection{Tuning-Remainder Model}

A USRP Hardware Driver (UHD) tune request is realized through an analog RF frequency and a residual
digital frequency correction,
$f_{\mathrm{req}} = f_{\mathrm{RF}} + s f_{\mathrm{DSP}}$, 
where the sign convention $s$ differs between the transmit and receive
paths~\cite{uhd_tuning_notes}. A mismatch between the realized TX and RX
digital corrections can therefore appear as a deterministic slow-time phase
rotation in the monostatic channel estimate.

An empirical model based on \SI{200}{\hertz}-quantized tuning remainders
reproduced the measured beat frequencies within \SI{0.002}{\hertz}. Although the internal origin of the apparent quantization was not
independently verified, the empirical model explains the observed
carrier-dependent rotation and provides a practical rule for selecting
operating frequencies with negligible measured TX--RX rotation.

\subsubsection{Digital and Hardware-Tuning Corrections}
A digital phase-reference correction can estimate the coupling phase at each slow-time sample and rotate the corresponding range profile to a common reference. This approach removes the deterministic rotation but requires a persistent high-SNR reference and propagates its phase-estimation error across the range profile. Thus, a strong nearby moving target may contaminate the phase-reference estimate, causing its motion-induced phase to be applied to the entire range profile.

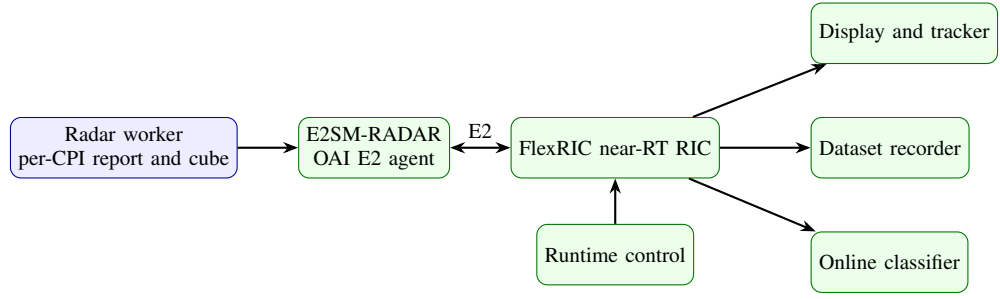
\begin{figure*}[t]
\centering
\begin{tikzpicture}[node distance=6mm and 8mm]
  \node[radar] (worker2) {Radar worker\\per-CPI report and cube};
  \node[oran,right=of worker2] (agent2) {E2SM-RADAR\\OAI E2 agent};
  \node[oran,right=of agent2] (ric2) {FlexRIC near-RT RIC};
  \node[oran,above right=7mm and 12mm of ric2] (display)
    {Display and tracker};
  \node[oran,right=12mm of ric2] (record) {Dataset recorder};
  \node[oran,below right=7mm and 12mm of ric2] (classify)
    {Online classifier};
  \node[oran,below=of ric2] (control) {Runtime control};

  \draw[link] (worker2) -- (agent2);
  \draw[bilink] (agent2) --
    node[above,font=\footnotesize]{E2} (ric2);
  \draw[link] (ric2) -- (display);
  \draw[link] (ric2) -- (record);
  \draw[link] (ric2) -- (classify);
  \draw[link] (control) -- (ric2);
\end{tikzpicture}
\caption{E2SM-RADAR processing and the implemented controller
applications. The gNB sends per-CPI detections and a compact slow-time slice
to the RIC. Runtime controls are transferred in the opposite direction and
applied by the radar worker at CPI boundaries.}
\label{fig:e2_arch}
\end{figure*}

A simpler mitigation is to select a carrier for which the measured residual
TX--RX rotation is negligible. Within band n78,
\SI{3.34848}{\giga\hertz} satisfied this criterion while preserving the 5G
cell numerology and bandwidth configuration. This carrier is used for all
subsequent measurements. The detailed coherence, clutter-suppression, and integration results before and after the carrier correction are reported in Section~\ref{sec:results}.

\section{O-RAN Sensing Service}
\label{sec:oran}

To expose the radar as a network-controlled sensing function, the
implementation follows the O-RAN architecture and its E2 interface toward the
near-real-time RAN Intelligent Controller~\cite{oran_foundational}. FlexRIC
~\cite{schmidt2021flexric} provides the controller and service-model
framework, while radar-specific information is carried through the custom
E2SM-RADAR service model described below. Figure~\ref{fig:e2_arch}
summarizes the E2 data path and the implemented controller-side applications.

\subsection{E2SM-RADAR}
The in-gNB radar initially produced detections only on the gNB host. To expose
the radar as a network function, a custom service model named E2SM-RADAR was
added to the FlexRIC and OAI E2 paths. E2SM-RADAR is specific to this
prototype and is not a standardized O-RAN Alliance E2 service model. It sends
one indication per completed CPI and supports control requests for runtime
parameters.

Each indication contains a fixed-size report with the timing, waveform,
calibration, and data-quality metadata required to interpret the CPI, together
with up to 64 detections.
Each detection reports the range and Doppler bins, physical range, radial
velocity, power, and estimated SNR. The occupied sensing bandwidth is
included explicitly because the achievable range resolution can vary with
the scheduler allocation.

At CPI completion, the radar worker publishes the fixed-size report and stages
the optional bulk payload under a separate lock. The E2 agent reads only these
published products, rather than the radar input ring, and encodes them into an
indication. This separation prevents E2 polling from contending with the radio
thread or the worker input path and allows the products to be consumed by
multiple controller applications without exposing real-time radio state.

The control message uses a bit mask to select any subset of the runtime parameters. The present controls include the CFAR scale, sounding interval, sounding-grid offset, minimum valid-sounding ratio, near-range blanking, zero-Doppler guard, phase-reference mode, and range-zero calibration. The gNB validates each request and returns an acceptance or rejection status
together with a reason. Parameters that require buffer reallocation, including CPI length and processing preset, remain restart-time settings.

Accepted controls are applied only at a CPI boundary. The worker finishes the current CPI, applies the new parameter set, clears its state, and starts the next CPI. This rule prevents one CPI from containing samples generated under different configurations.

\subsection{Compact Slow-Time Payload}
In addition to the fixed-size report, an indication may carry an optional
complex slow-time slice over a small range window before clutter cancellation.
A detection list is sufficient for tracking but does not preserve
micro-Doppler energy below the CFAR threshold; the slice retains this
information for display, recording, and classification. At the operating point in Table~\ref{tab:operating_point}, the payload is
5\,264~bytes per CPI, corresponding to
\SI{263}{\kilo\bit\per\second}. This is much smaller
than a continuous IQ stream.

The slow-time slice is extracted before mean removal because independently
removing the mean from each CPI would introduce boundary discontinuities in a
controller-side spectrogram. The samples remain complex so that the controller
can select its own window, overlap, range gate, and clutter treatment. A
validity mask is sent with the slow-time slice because an invalid column would
otherwise appear as broad Doppler energy.

Table~\ref{tab:e2_products} summarizes the data products and the purpose of transporting each over E2. Raw IQ and the full range--Doppler map remain in the gNB. The controller receives data that supports monitoring and decisions without moving the highest-rate signal.

\begin{table}[ht]
\centering
\caption{E2SM-RADAR Data Products}
\label{tab:e2_products}
\footnotesize
\begin{tabular}{L{2.25cm}L{2.0cm}L{3.1cm}}
\toprule
Product & Rate at the selected point & Purpose\\
\midrule
CPI metadata & one record per \SI{160}{\milli\second} & Timing, calibration, bandwidth, and health\\
Detection list & 0--64 records per CPI & Tracking, alarms, and range--velocity display\\
Slow-time slice & \SI{5264}{\byte} per CPI & Micro-Doppler, recording, and classification\\
Validity mask & one bit per slow-time column & Missing-column handling\\
Control request & event-driven & CFAR, cadence, range gate, and calibration\\
\bottomrule
\end{tabular}
\end{table}

\subsection{Controller Applications}
\label{sec:e2_service}

Four controller-side xApps were implemented. The configuration xApp monitors
the radar status and updates runtime parameters. The display xApp combines
detection reports and slow-time slices to generate range--time and
velocity--time views with overlaid CFAR detections and to maintain a
constant-velocity track. The recording xApp stores labeled measurements
together with their radar configuration and data-quality counters. The
classification xApp extracts features over a sliding \SI{2}{\second} window
and applies a linear model trained offline.

The display and classification xApps process complex slow-time slices over
their respective analysis windows and apply clutter removal within each
window. CPI timestamps are used to account for missing CPIs, while the
validity mask identifies invalid soundings when constructing the slow-time
axis. This avoids estimating a single background over the complete recording.
The classification xApp uses the same feature definitions and model
parameters during offline training and online inference.

\section{Experimental Results}
\label{sec:results}

\subsection{Hardware-Tuning Correction and Coherent Integration}
Table~\ref{tab:phase_result} summarizes the impact of carrier selection without digital phase compensation. For the X300, the measured coupling rotation at \SI{3.32976}{\giga\hertz} was \SI{28.5714}{\hertz}, which reduced to \SI{-0.0012}{\hertz} when the carrier frequency was shifted to \SI{3.34848}{\giga\hertz}. Concurrently, the mean-removal clutter-suppression metric improved from \SI{-16.4}{\decibel}, indicating clutter enhancement, to \SI{38.0}{\decibel}, and the coherent-integration loss at  $M=64$ decreased from \SI{19.81}{\decibel} to \SI{0.27}{\decibel}. As a baseline, the N310 exhibited negligible rotation across all tested carriers, serving as an independent reference.

\begin{table}[ht]
\centering
\caption{Measured Effect of the Carrier-Grid Correction Without Digital Phase Compensation}
\label{tab:phase_result}
\footnotesize
\setlength{\tabcolsep}{2.5pt}
\begin{tabular}{L{3.05cm}ccc}
\toprule
Metric &
\shortstack{X300\\3.32976 GHz} &
\shortstack{X300\\3.34848 GHz} &
\shortstack{N310\\3.32976 GHz} \\
\midrule
TX--RX rotation (\si{\hertz})
    & 28.5714 & $-0.0012$ & $-0.0002$ \\
Apparent velocity (\si{\meter\per\second})
    & 1.286 & $\approx 0$ & $ 0$ \\
Clutter suppression (\si{\decibel})
    & $-16.4$ & 38.0 & 43.7 \\
Integration loss, $M=64$ (\si{\decibel})
    & 19.81 & 0.27 & 0.06 \\
\bottomrule
\end{tabular}
\end{table}

The corrected-carrier measurements also quantify the available coherent
integration interval. Using $M=8$ as a reference and comparing integrations
formed from the same sequence of consecutive valid soundings, we found that the static
coupling follows the ideal coherent-integration law to within
\SI{0.01}{\decibel} for all tested lengths up to $M=128$ (corresponding to
\SI{320}{\milli\second}). At the deployed $M=64$, the measured
$-3$\,dB Doppler width is \SI{0.410}{\meter\per\second}, close to the
\SI{0.403}{\meter\per\second} predicted for the applied Hann weighting.
These results indicate that radio coherence does not limit the deployed
\SI{160}{\milli\second} CPI; the practical limit is instead set by target
nonstationarity. By contrast, at the uncorrected carrier, the measured
TX--RX phase rotation completes one cycle approximately every
\SI{35}{\milli\second}, severely limiting coherent integration.

\subsection{Measurement Method}
For the experimental measurements, the hardware configuration is detailed in
Table~\ref{tab:operating_point}.  Six scenarios were set up for raw-IQ data recordings: 1) an empty
room, 2) stationary corner reflectors at \SI{3}{\meter}, 3) \SI{6}{\meter}, 4)
\SI{9}{\meter}, 5) a person walking over approximately \SIrange{3}{12}{\meter},
and 6) the same walk while carrying a metal sheet. The only human participant was one of the authors, who
provided informed consent for the recordings and their use in this study.

Each evaluation unit is a non-overlapping \SI{160}{\milli\second} CPI. For moving-target experiments, a detection is considered to be correct if it lies within
$\pm1$ range bin (\SI{2.44}{\meter}) and $\pm2$ Doppler bins
(\SI{0.56}{\meter\per\second}) of the reference cell. CPIs with
$|v|<\SI{0.6}{\meter\per\second}$ are excluded because they fall inside the
MTI notch, leaving 42 evaluable CPIs per walk.

The walking recordings are used for processing-gain, detection-sensitivity,
and range-consistency measurements. The stationary-reflector recordings are
used only for false-alarm evaluation: direct coupling and room clutter remain
strong over their range cells, and changes in the complex background between
recordings prevent reliable subtraction of the empty-room response.
Consequently, they are not used for detection-probability or
radar-cross-section (RCS) measurements.

\subsection{Processing Gain}
\label{sec:gain}

For coherent symbol-domain processing, the ideal SNR gain scales with the
number of coherently processed resource elements~\cite{SW_performance_verification}.
With $N_{\mathrm{occ}}=1943$ occupied subcarriers,
$L_{\mathrm{s}}=14$ OFDM symbols per sounded slot, and $M=64$ slow-time
samples, the ideal processing gain is
$10\log_{10}(N_{\mathrm{occ}}L_{\mathrm{s}}M)=\SI{62.41}{\decibel}$.
Accounting for the range and Doppler windows, low-amplitude masking, and missing
soundings gives the predicted gain of \SI{57.56}{\decibel} in
Table~\ref{tab:gain_budget}.

\begin{table}[ht]
\centering
\caption{Processing-Gain Budget for the Selected Configuration}
\label{tab:gain_budget}
\footnotesize
\begin{tabular}{L{4.4cm}rr}
\toprule
Term & Value [dB] & Cumulative [dB]\\
\midrule
$10\log_{10}(N_{\mathrm{occ}}L_{\mathrm{s}}M)$ & & 62.41\\
Range window & $-1.76$ & 60.65\\
Doppler window & $-1.83$ & 58.82\\
Low-amplitude mask & $-0.27$ & 58.55\\
Missing soundings & $-0.98$ & 57.56\\
\midrule
\textbf{Measured} & & \textbf{57.62}\\
Residual (measured $-$ predicted) & & $+0.06$\\
\bottomrule
\end{tabular}
\end{table}

The gain was measured from the recorded walking data by injecting complex
Gaussian noise and comparing the input SNR with the post-integration SNR in
noise-limited CPIs. Across 68 measurements, the realized gain was
\SI{57.62}{\decibel}, only \SI{0.06}{\decibel} above the prediction, with a
standard deviation of \SI{1.30}{\decibel}. The output SNR followed the
injected level with a slope of 0.990.

\subsection{Link Budget and Power-Limitation Characterization}
\label{sec:link_budget}
For a monostatic point target, the free-space radar equation can be written
in logarithmic form as
\begin{equation}
\begin{split}
P_r ={}& P_{\mathrm{TX}} + G_{\mathrm{TX}} + G_{\mathrm{RX}}
-2L_{\mathrm{feed}}
+20\log_{10}\lambda  \\
&+10\log_{10}\sigma
-30\log_{10}(4\pi)
-40\log_{10}R,
\end{split}
\label{eq:link_budget}
\end{equation}
where $P_{\mathrm{TX}}$ is referenced to the USRP TX port, $\sigma$ is the
target radar cross section, and $R$ is the target range. With
$\lambda=\SI{89.5}{\milli\meter}$, transmit and receive antenna gains of
\SI{6}{\dBi}, $\sigma=\SI{1}{\meter\squared}$, and an estimated feed loss of \SI{1.6}{\decibel} per path,

\begin{equation}
P_r[\mathrm{dBm}] = P_{\mathrm{TX}}[\mathrm{dBm}]
-45.14-40\log_{10}R \label{eq:link_budget_reduced}.
\end{equation}

The transmit power was measured directly at the X300 TX SMA using a spectrum
analyzer in channel-power mode over the \SI{58}{\mega\hertz} occupied band.
After accounting for the external \SI{30}{\decibel} attenuator and the
\SI{74.3}{\percent} downlink duty cycle, the on-burst power was
\SI{-0.7}{\dBm}. With the estimated feed loss, this corresponds to
approximately \SI{-2.3}{\dBm} at the antenna port and an effective isotropic radiated power of
approximately \SI{3.7}{\dBm}. The OAI-declared downlink reference-signal
power is therefore not used as an RF calibration quantity.

Using the measured processing gain of \SI{57.62}{\decibel} from
Table~\ref{tab:gain_budget}, together with an estimated receiver noise figure
of \SI{7}{\decibel}, gives a thermal-noise power of approximately
\SI{-89.3}{\dBm} over \SI{58.29}{\mega\hertz}. For
$\sigma=\SI{1}{\meter\squared}$, the resulting post-processing SNR is
approximately \SI{66.5}{\decibel} at \SI{7.3}{\meter}, and a
\SI{13}{\decibel} detection threshold gives a nominal noise-limited range under the stated assumptions of
approximately \SI{159}{\meter}. These values provide a free-space reference
rather than a prediction of the indoor operating range because target RCS,
receiver noise figure, antenna gain, and multipath are not independently
calibrated.

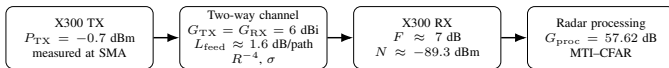
\begin{figure}[ht]
\centering
\resizebox{\columnwidth}{!}{%
\begin{tikzpicture}[
    >=Latex,
    font=\scriptsize,
    block/.style={
        draw,
        rounded corners,
        align=center,
        minimum height=12mm,
        text width=27mm,
        inner sep=2pt
    },
    arrow/.style={->,thick}
]
\node[block] (tx) {
    X300 TX\\
    $P_{\mathrm{TX}}=-0.7$ dBm\\
    measured at SMA
};

\node[block, right=5mm of tx] (channel) {
    Two-way channel\\
    $G_{\mathrm{TX}}=G_{\mathrm{RX}}=6$ dBi\\
    $L_{\mathrm{feed}}\approx1.6$ dB/path\\
    $R^{-4}$, $\sigma$
};

\node[block, right=5mm of channel] (rx) {
    X300 RX\\
    $F\approx7$ dB\\
    $N\approx-89.3$ dBm
};

\node[block, right=5mm of rx] (proc) {
    Radar processing\\
    $G_{\mathrm{proc}}=57.62$ dB\\
    MTI--CFAR
};

\draw[arrow] (tx) -- (channel);
\draw[arrow] (channel) -- (rx);
\draw[arrow] (rx) -- (proc);
\end{tikzpicture}%
}
\caption{Measured and estimated terms used in the radar link budget. The
transmit power and processing gain are measured quantities, whereas the feed
loss, receiver noise figure, antenna gains, and target RCS retain the stated
model assumptions.}
\label{fig:link_budget}
\end{figure}

The measured system is nevertheless not power-limited over the tested indoor
range. In a nine-point transmit-gain sweep, the dominant static return tracked
the commanded TX-gain reduction to within \SI{0.3}{\decibel}. The post-MTI floor decreased with transmit power initially and then saturated at a level \SI{22.1}{\decibel} below its full-power value. Reducing the transmit power by
\SI{20}{\decibel} changed the wall-to-floor dynamic range from
\SI{73.19}{\decibel} to \SI{72.32}{\decibel}, a loss of only
\SI{0.9}{\decibel}; a \SI{15}{\decibel} reduction increased the dynamic
range to \SI{74.73}{\decibel}. An independent RX-gain experiment increased the analog receive gain by \SI{12.2}{\decibel} at deep TX backoff but improved the wall-to-floor dynamic range by only \SI{1.04}{\decibel}, indicating that the limiting floor is dominated by receiver noise rather than analog-to-digital converter quantization. Thus, at the nominal transmit setting, approximately
\SI{22}{\decibel} separates the clutter-limited working floor from the
receiver-noise floor. Increasing transmit power therefore provides little
benefit; improved isolation and suppression of strong static returns are more
relevant to extending the useful sensing range.

\subsection{Radar Detection Performance}
\begin{figure}[ht]
\centering
\includegraphics[width=\columnwidth]{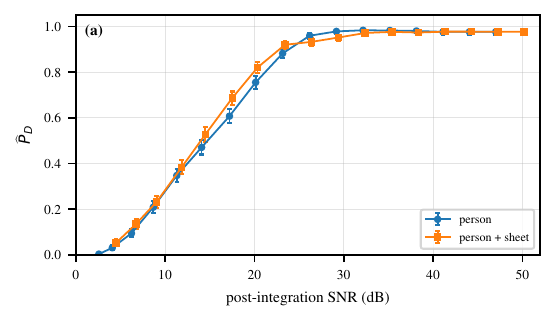}\\[-1mm]
\includegraphics[width=\columnwidth]{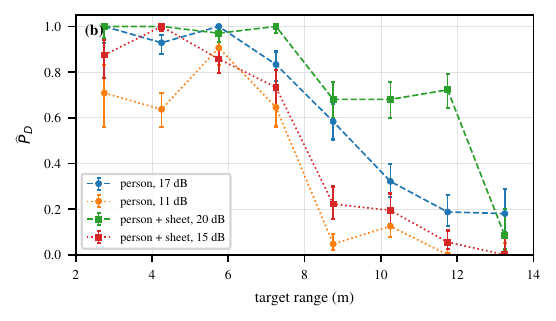}
\caption{Detector sensitivity at the deployed OS-CFAR scale of 15.
(a)~Conditional detection fraction versus detector-relative post-integration
SNR. (b)~Conditional detection fraction versus range at two added-noise
levels per target, grouped in \SI{1.5}{\meter} bins. Results use the
42 evaluable CPIs per recorded trajectory defined above.}
\label{fig:detection_results}
\end{figure}

Detector sensitivity was evaluated by adding complex Gaussian noise to each
of the 42 valuable CPIs from each walking recording and re-running the complete
processing and detection chain. Each added-noise level uses 24 independent
noise realizations per recorded CPI. These realizations provide repeated
noise conditions for the same target snapshots rather than independent target
trials, so binomial confidence intervals are not reported. The conditional
detection fraction is
\[
\widehat P_{D,\mathrm{cond}}
=
\frac{N_D}{N_{\mathrm{eval}}},
\]
and applies only to the recorded trajectories outside the MTI notch.
Post-integration SNR is measured relative to the local background within the
searched region and is therefore a detector-relative quantity rather than a
calibrated echo SNR. 
Figure~\ref{fig:detection_results} characterizes detector sensitivity at the
deployed OS-CFAR scale of 15.

Over the recordings used for false-alarm evaluation, the empirical false-alarm
rate was $\widehat P_{FA}=6.9\times10^{-4}$. For both recorded moving-target
trajectories, the conditional detection fraction exceeded 0.9 above
approximately \SI{25}{\decibel} detector-relative post-integration SNR and was
0.976 at the original recorded noise level. With added noise, detection
decreased with range, although this dependence can only be assessed coarsely
because the \SIrange{3}{12}{\meter} trajectory spans only a few
range-resolution cells.

The remaining false alarms were dominated by a small number of persistent
range--Doppler cells rather than by thermal noise: \SI{68}{\percent} occurred
in only five cells. Suppressing repeatedly triggered cells reduced the
false-alarm rate more effectively than increasing the CFAR threshold alone,
indicating that persistent clutter locations are the main limitation of the
deployed detector.

\subsection{Sensing Bandwidth and Range Resolution}
\label{sec:bwsweep}

The occupied sensing bandwidth of a scheduled system follows the downlink
allocation. To characterize this dependence separately from the traffic that
produces it, the walking recording was reprocessed with the occupied band
restricted to 162, 106, 51, and 24 physical resource blocks (PRBs). The range
window was refitted to each band edge, and the complete detection chain was
rerun.
Table~\ref{tab:bw_sweep} reports the result.

\begin{table}[ht] \centering \caption{Measured Effect of the Occupied Sensing Bandwidth} \label{tab:bw_sweep} \footnotesize \setlength{\tabcolsep}{2.7pt} \begin{tabular}{rrrrrr} \toprule PRBs & $B$ & $\Delta R$ & Measured $-3$-dB extent &
$\Delta\mathrm{SNR}$ & $\widehat P_{D,\mathrm{cond}}$\\ & [MHz] & [m] & [m] & [dB] & \\ \midrule 162 & 58.29 & 2.57 & 4.86 & $0.0$ & 0.958\\ 106 & 38.1 & 3.93 & 6.41 & $-1.8$ & 0.949\\ 51 & 18.3 & 8.18 & 12.13 & $-6.3$ & 0.808\\ 24 & 8.6 & 17.41 & 25.17 & $-12.4$ & 0.439\\ \bottomrule \end{tabular} \end{table}

The measured $-3$-dB range extent of the target follows the Hann-weighted
prediction, $1.44c_0/(2B)$, at the narrower bandwidths. At the widest
bandwidth, the resolution cell becomes smaller than the walking subject, and
the measured extent is therefore limited by the target extent rather than by
the instrumental resolution. At a fixed added-noise level, the conditional detection fraction falls from 0.958 to 0.439 over the same bandwidth reduction. Peak SNR falls by \SI{12.4}{\decibel} in total, of which
\SI{8.3}{\decibel} is the reduced number of coherently combined resource
elements; the remainder is the local background rising as the resolution cell
widens, since this SNR is referenced to the median of the searched region.

\begin{figure*}[!t]
\centering
\includegraphics[width=0.485\textwidth]{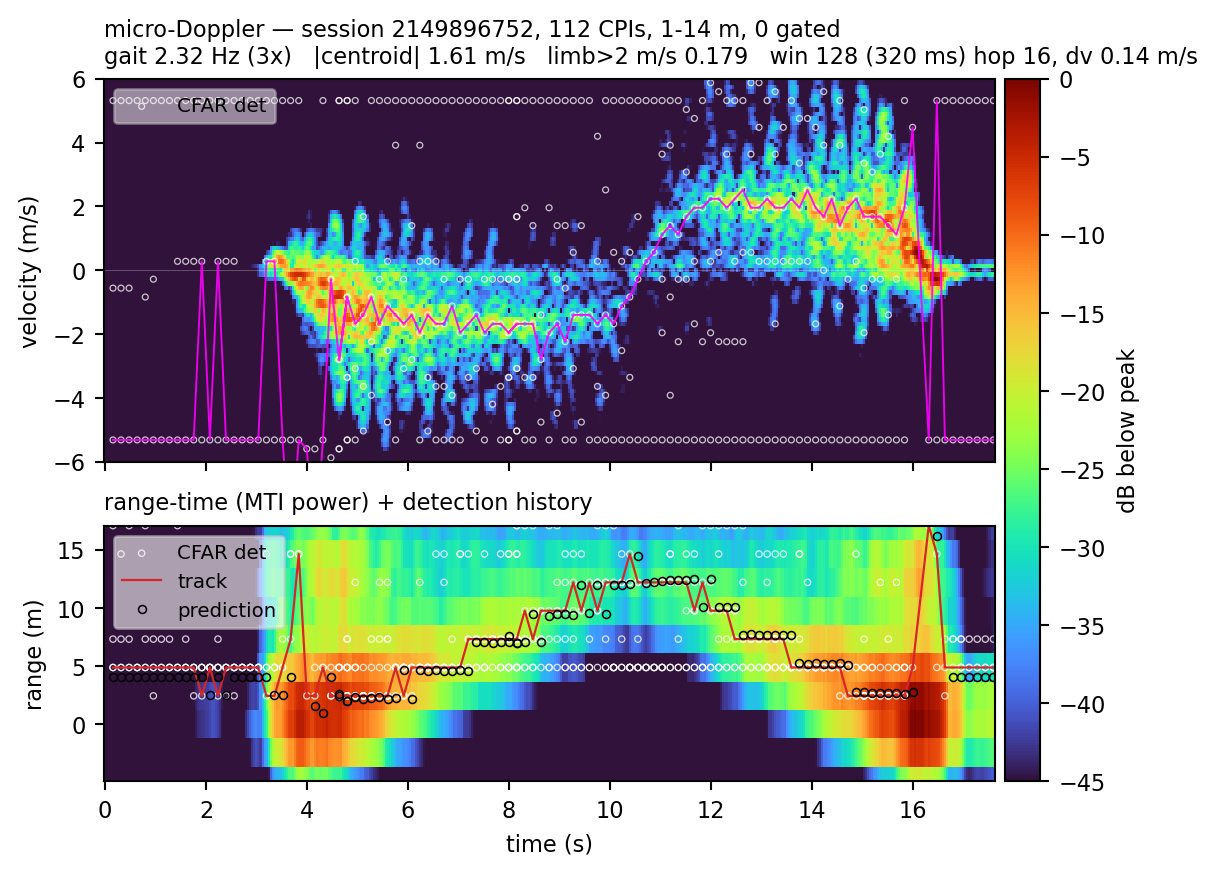}\hfill
\includegraphics[width=0.485\textwidth]{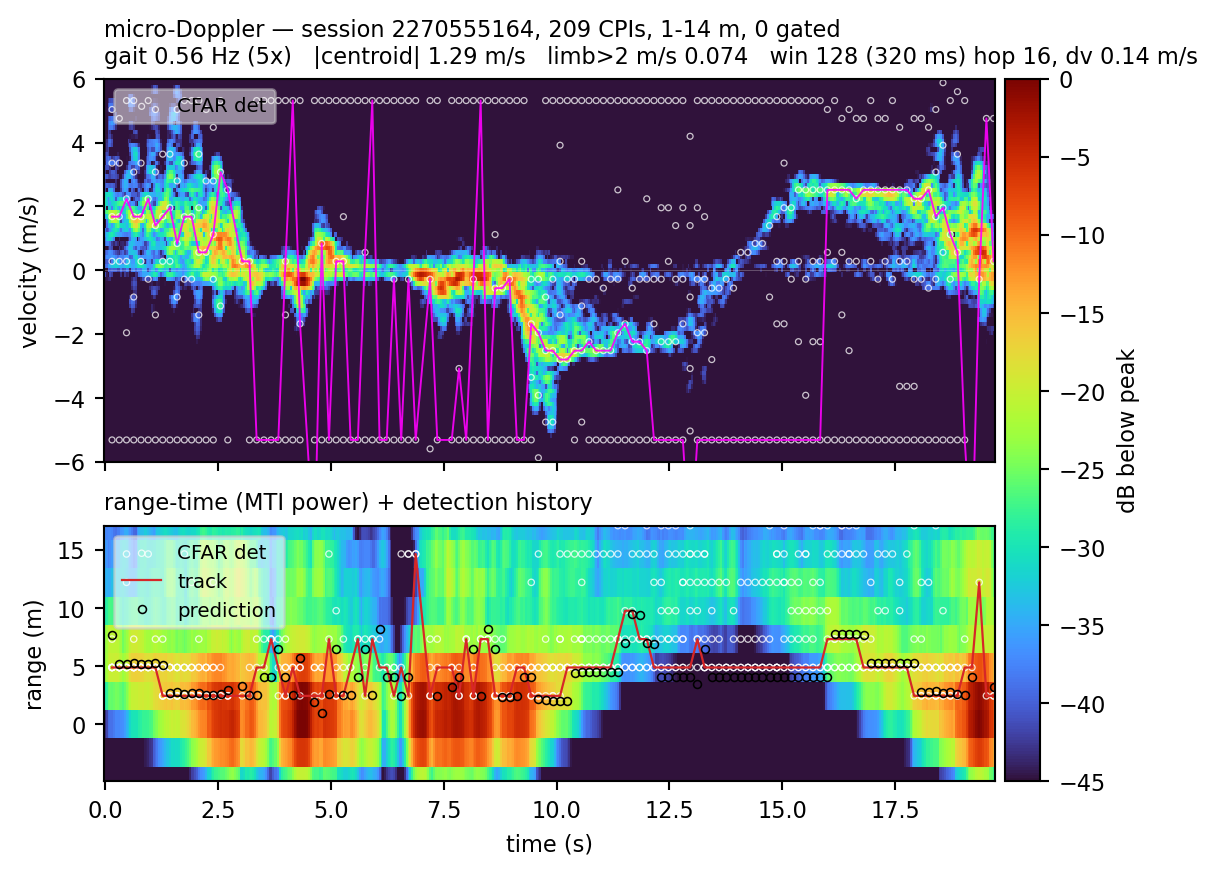}
\caption{Controller-side micro-Doppler and range--time output reconstructed
from the compact E2 payload. (a)~Walking subject, showing the Doppler-sign
and range-slope reversal at the turnaround and limb components around the
torso motion. (b)~Wheeled-chair motion along the same radial path, showing a
different micro-Doppler structure at a similar radial speed.}
\label{fig:live_md_examples}
\end{figure*}

\subsection{Real-Time Execution and Queue Stability}
\label{sec:rt_exec}
The background-worker architecture in Section~\ref{sec:implementation}
removes radar processing from the slot-critical gNB thread. For the deployed
configuration to operate in real time, the worker must sustain the incoming
sounding rate without accumulating backlog or dropping input data. The
measurements below verify this condition from the per-sounding and per-CPI
execution times, input-ring occupancy, and drop counter.

All timing measurements were obtained on the host summarized in
Table~\ref{tab:host_platform}: a 12-core Intel Xeon W-3323 with a
\SI{3.5}{\giga\hertz} base clock, running a \code{PREEMPT\_RT} Linux kernel,
with simultaneous multithreading disabled and the USRP X300 connected through
a \SI{10}{\giga\bit\per\second} Ethernet interface. The radar worker executes
as an ordinary background thread on the same host as the gNB and is not pinned
to a dedicated core. No CPU isolation, real-time scheduling priority, or
frequency locking was applied beyond the standard \code{PREEMPT\_RT}
configuration. The timing results reported below therefore characterize an
untuned general-purpose deployment rather than a platform optimized
specifically for deterministic radar execution and provide a conservative
empirical estimate of the available processing margin.

\begin{table}[ht]
\centering
\caption{gNB Host Platform Used for Real-Time Measurements}
\label{tab:host_platform}
\footnotesize
\setlength{\tabcolsep}{3pt}
\begin{tabular}{L{3.4cm}l}
\toprule
Component & Specification\\
\midrule
Platform             & Supermicro SYS-540A-TR\\
Processor            & Intel Xeon W-3323\\
Physical cores       & 12 (SMT disabled)\\
Base clock           & \SI{3.50}{\giga\hertz}\\
Maximum clock        & \SI{3.90}{\giga\hertz}\\
Last-level cache     & \SI{21}{\mebi\byte} shared L3\\
Vector extensions    & AVX-512 (F, DQ, BW, VL, VNNI)\\
System memory        & \SI{128}{\gibi\byte}, single NUMA domain\\
Operating system     & Ubuntu \num{24.04} LTS\\
Kernel               & Linux \num{6.8}, \code{PREEMPT\_RT}\\
CPU frequency policy & \code{ondemand}, turbo enabled\\
Core isolation       & None (\code{isolcpus} and \code{nohz\_full} not used)\\
Host--SDR interface  & Intel X550 \SI{10}{\giga\bit\per\second} Ethernet\\
\bottomrule
\end{tabular}
\end{table}

Execution time was measured inside the worker 
during a separate instrumented run containing 1145
CPIs and 6425 target detections. The timing histograms contain 1088 complete
CPIs and their 69\,632 constituent soundings. Table~\ref{tab:rt_exec}
summarizes the execution-time and queue measurements.

\begin{table}[ht]
\centering
\caption{Radar-Worker Real-Time Execution Measurements}
\label{tab:rt_exec}
\footnotesize
\setlength{\tabcolsep}{3pt}
\begin{tabular}{L{5.0cm}r}
\toprule
Metric & Measured\\
\midrule
\multicolumn{2}{l}{\emph{Per-sounding stage ($n=69\,632$)}}\\
\quad Median                    & \SI{0.314}{\milli\second}\\
\quad 95th percentile           & \SI{0.403}{\milli\second}\\
\quad 99th percentile           & \SI{0.498}{\milli\second}\\
\quad Maximum                   & \SI{0.619}{\milli\second}\\
\midrule
\multicolumn{2}{l}{\emph{Per-CPI stage ($n=1088$)}}\\
\quad Median                    & \SI{52.22}{\milli\second}\\
\quad 95th percentile           & \SI{53.08}{\milli\second}\\
\quad 99th percentile           & \SI{53.40}{\milli\second}\\
\quad Maximum                   & \SI{53.76}{\milli\second}\\
\midrule
CPI utilization, median-stage estimate       & \SI{45.2}{\percent}\\
CPI utilization, conservative bound   & \SI{58.4}{\percent}\\
Input-ring capacity                       & 64 soundings\\
Maximum observed ring occupancy           & 22 soundings\\
Soundings dropped                         & 0\\
\bottomrule
\end{tabular}
\end{table}

The measured execution times confirm real-time operation at the deployed
configuration. Using the median stage times, the combined per-sounding and
per-CPI workload is \SI{72.32}{\milli\second} per
\SI{160}{\milli\second} CPI, corresponding to \SI{45.2}{\percent}
utilization. Even the conservative bound formed by assigning the maximum
observed per-sounding time to all 64 soundings and combining it with the
maximum per-CPI time is only \SI{93.38}{\milli\second}, or
\SI{58.4}{\percent} of the CPI interval. The maximum per-CPI execution time
of \SI{53.76}{\milli\second} predicts $53.76/2.5=21.5$ soundings arriving
while CPI processing is active, closely matching the maximum measured ring
occupancy of 22 soundings. The queue therefore reflects the expected
CPI-processing burst rather than progressive backlog, and no soundings were
dropped during the instrumented run.

\subsection{Live E2 Transport and Stream Integrity}
\label{sec:live_session}

After establishing real-time radar execution in
Section~\ref{sec:rt_exec}, the complete gNB--RIC chain was evaluated to
determine whether sensing products could be delivered continuously to the
near-RT RIC without loss or excessive transport overhead. Six live runs
produced 2591 CPI reports over \SI{416}{\second}.
Table~\ref{tab:live_e2health} summarizes the resulting transport and
stream-integrity measurements.

\begin{table}[ht]
\centering
\caption{E2 Transport and Stream Integrity During the Live Session}
\label{tab:live_e2health}
\footnotesize
\begin{tabular}{L{4.6cm}r}
\toprule
Quantity & Measured\\
\midrule
Session duration                 & \SI{416}{\second}\\
CPI reports delivered            & 2591\\
Slow-time samples delivered      & 165\,824\\
Invalid samples                  & 8 (\SI{0.005}{\percent})\\
Radar-ring drops                 & 0\\
Noncontiguous CPI counter steps  & 0\\
CPI-publication-to-xApp delay    & \SIrange{0.3}{0.5}{\milli\second}\\
Payload per CPI                  & \SI{5264}{\byte}\\
Sustained payload rate           & \SI{263}{\kilo\bit\per\second}\\
\bottomrule
\end{tabular}
\end{table}

The contiguous CPI sequence and zero ring-drop count show that no complete CPI
reports were lost during the recorded runs, while the
\SI{0.005}{\percent} invalid-sample fraction indicates that the slow-time
product remained essentially complete. The sub-millisecond delivery delay and
\SI{263}{\kilo\bit\per\second} payload rate further show that E2 transport
does not form a bottleneck for the deployed sensing configuration.

\subsection{Controller-Side Micro-Doppler and Tracking}
\label{sec:live_md}

Figure~\ref{fig:live_md_examples} shows controller-side micro-Doppler and
range--time products reconstructed from the compact E2 payload. The
velocity--time spectrogram uses a 128-sample short-time Fourier transform
with a 16-sample hop, corresponding to a \SI{320}{\milli\second} window and
a \SI{0.14}{\meter\per\second} velocity cell. The range--time view shows MTI
power together with the CFAR detections and tracker output.

For the walking example, the torso component lies near
$\pm\SI{1.5}{\meter\per\second}$, while limb components extend to
approximately $\pm\SI{4.5}{\meter\per\second}$. The subject moves from
approximately \SI{2.5}{\meter} to \SI{12.5}{\meter}, turns, and returns.
The simultaneous reversal of the Doppler sign and range slope provides a
consistency check between the slow-time payload and the independently
transported detection list. Figure~\ref{fig:live_md_examples}(b) shows
wheeled-chair motion over the same path, with a visibly different
micro-Doppler structure at a similar radial speed.

Across the live session, the tracker produced 2581 one-step residuals over
2607 updates. The residual RMS was \SI{1.44}{\meter}, with
\SI{85.7}{\percent} of residuals within one \SI{2.44}{\meter} range bin;
20 reacquisitions occurred, mainly near turnarounds. For comparison, two independently quantized range measurements with uniform
errors over one \SI{2.44}{\meter} bin have an expected RMS difference of
approximately \SI{1.00}{\meter}. Thus, much of the observed residual can be
attributed to the available range resolution.

Range differencing is therefore poorly suited to velocity estimation at this
resolution. A one-bin change over one \SI{160}{\milli\second} update
corresponds to \SI{15.2}{\meter\per\second}, exceeding the radar's
\SI{8.95}{\meter\per\second} unambiguous velocity. Consequently,
\SI{99.6}{\percent} of range-slope velocity estimates exceeded this limit,
compared with only \SI{1.6}{\percent} of Doppler estimates. The Doppler
centroid is therefore used as the velocity measurement for tracking.

\subsection{Micro-Doppler Classification}
\label{sec:classification}

The classification experiment tests whether the compact E2 payload retains
enough target-dependent structure for controller-side inference. It serves as
an end-to-end demonstration of the radar, E2 transport, recording, training,
and xApp inference path using a lightweight classifier.

The dataset contains 17 cases: nine walking and eight wheeled-chair
recordings, collected with one subject, one room, and one radar geometry.
Each sample is a \SI{2}{\second} window of the slow-time stream, corresponding
to 50 short-time Fourier transform (STFT) frames at a
\SI{40}{\milli\second} hop. Windows with a moving-to-static power ratio below
\SI{6}{\decibel} are discarded, leaving 191 windows: 104 walking and 87
wheeled-chair.

Motion direction and speed were selected to make radial speed similar between
the two classes. The median speeds are \SI{1.43}{\meter\per\second} for
walking and \SI{1.57}{\meter\per\second} for the wheeled chair, and
\SI{91.6}{\percent} of all windows lie within their common
\SIrange{0.50}{2.45}{\meter\per\second} support. The speed distributions have
a Bhattacharyya coefficient of 0.78 using
\SI{0.2}{\meter\per\second} bins and an overlapping coefficient of 0.67 from
a Gaussian kernel-density estimate. Thus, radial speed alone provides little
transferable class information.

An $\ell_2$-regularized binary logistic-regression classifier is trained on
11 micro-Doppler features describing spectral shape, limb energy, gait rate,
entropy, torso dominance, and related motion characteristics. A twelfth
candidate feature, target radial speed, is excluded from the deployed model
so that classification depends primarily on micro-Doppler structure. Features
are standardized using training-fold statistics only.

Validation uses leave-one-take-out cross-validation, with each fold holding
out one complete take so that no window from that recording appears in the
training set. Over 191 windows, the speed-excluded classifier achieves
\SI{84.8}{\percent} per-window accuracy and
\SI{90.8}{\percent} mean per-take accuracy, while majority voting correctly
classifies all 17 held-out takes. Retaining target speed increases the
corresponding accuracies only modestly, to \SI{88.5}{\percent} and
\SI{92.9}{\percent}. The majority-class baseline is
\SI{54.5}{\percent}; within the matched
\SIrange{0.6}{1.0}{\meter\per\second} speed interval, the speed-excluded
classifier achieves \SI{58.1}{\percent}, indicating that class separation is
weaker when radial-speed differences are minimized.

Table~\ref{tab:classification} summarizes the validation results.

\begin{table}[ht]
\centering
\caption{Walking Versus Wheeled-Chair Classification}
\label{tab:classification}
\footnotesize
\begin{tabular}{lcc}
\toprule
Model & Per-window & Per-take\\
\midrule
Majority-class baseline
    & \SI{54.5}{\percent} & --\\
Speed-only baseline
    & \SI{9.4}{\percent} & \SI{25.0}{\percent}\\
Logistic regression, speed excluded
    & \SI{84.8}{\percent} & \SI{90.8}{\percent}\\
Logistic regression, all features
    & \SI{88.5}{\percent} & \SI{92.9}{\percent}\\
\midrule
Per-take majority vote, speed excluded
    & -- & 17/17\\
\bottomrule
\end{tabular}
\end{table}

The same speed-excluded model was deployed in the controller xApp without
retraining. Figure~\ref{fig:live_md_examples} shows different micro-Doppler
structure for walking and wheeled-chair motion at similar radial speeds.
During the live session, 82 of 91 window-level predictions agreed with the
corresponding activity label, giving an agreement rate of
\SI{90.1}{\percent}.

\subsection{Multitarget Separation}
\label{sec:two_targets}

The preceding measurements consider a single moving target. To test
simultaneous-target separation, a wheeled chair was released toward the radar
from approximately \SI{8}{\meter} while the operator walked away along the
same radial path. Raw slot-level IQ was recorded for
\SI{10.9}{\second} and reprocessed with the same division, range, MTI, and
Doppler chain used by the online radar. This experiment used a
\SI{5}{\milli\second} slow-time interval and $M=64$, giving a
\SI{320}{\milli\second} processing window, a
\SI{0.14}{\meter\per\second} velocity-bin spacing, and an unambiguous
velocity of $\pm\SI{4.48}{\meter\per\second}$. The range-bin spacing remained
\SI{2.44}{\meter}.

\begin{figure}[ht]
  \centering
  \includegraphics[width=\columnwidth]{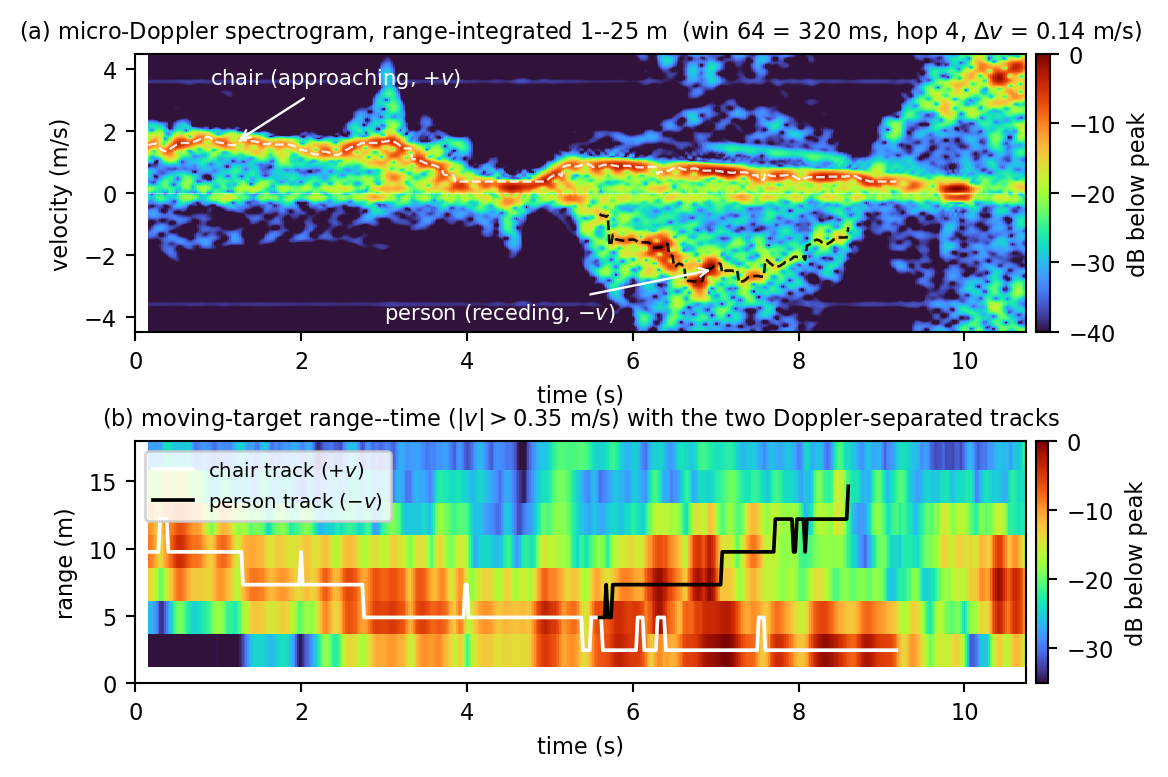}
  \caption{Evolution of two simultaneously moving targets.
  (a)~Range-integrated micro-Doppler spectrogram; the approaching chair
  occupies the positive Doppler half-plane and the retreating operator the
  negative half-plane. (b)~Moving-target range--time map showing the two
  target trajectories.}
  \label{fig:two_targets}
\end{figure}

Figure~\ref{fig:two_targets} shows the target evolution in Doppler and range.
The two targets remain in opposite Doppler half-planes while moving. Under the adopted sign convention, the chair has a mean Doppler velocity of
$+\SI{0.94}{\meter\per\second}$ and a range rate of
$-\SI{0.91}{\meter\per\second}$, whereas the operator has a mean Doppler
velocity of $-\SI{2.02}{\meter\per\second}$ and a range rate of
$+\SI{2.41}{\meter\per\second}$. The Doppler signs are therefore
consistent with the corresponding range motion.

\begin{figure}[ht]
  \centering
  \includegraphics[width=\columnwidth]{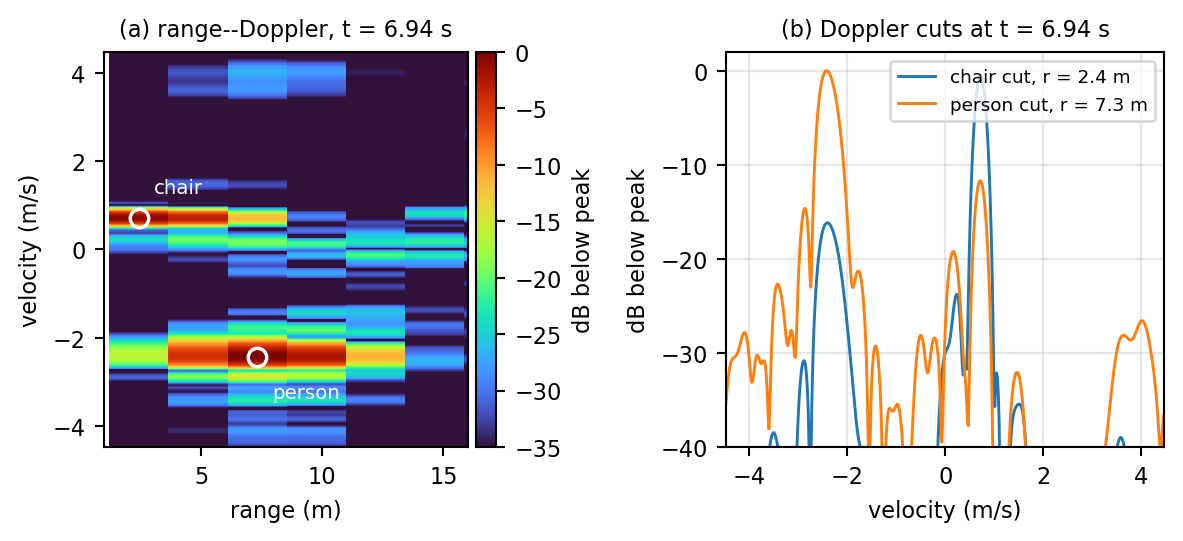}
  \caption{Two-target snapshot at $t=\SI{6.94}{\second}$.
  (a)~Range--Doppler map showing both targets resolved in range and Doppler.
  (b)~Doppler cuts at the two target ranges, showing the narrower chair
  return and the broader micro-Doppler extent of the walking subject.}
  \label{fig:two_targets_snapshot}
\end{figure}

During the approximately \SI{3}{\second} interval in which both targets are
moving, their mean separation is \SI{6.5}{\meter} in range and
\SI{2.7}{\meter\per\second} in Doppler, corresponding to approximately
2.6 range bins and 19 velocity bins. Both targets remain more than
\SI{20}{\decibel} above the local noise floor in all 151 overlapping
processed windows. Figure~\ref{fig:two_targets_snapshot} shows a representative
range--Doppler snapshot in which both targets are resolved along both
dimensions. Thus, the same processing chain separates the two moving targets
simultaneously in range and Doppler.

The Doppler spread also differs between the two targets. At
\SI{10}{\decibel} below the peak, the chair occupies approximately
\SI{0.50}{\meter\per\second}, compared with
\SI{1.77}{\meter\per\second} for the walking subject. The broader human
return is consistent with additional limb motion and provides the type of
micro-Doppler structure used by the classifier in
Section~\ref{sec:classification}. Because the targets do not occupy the same
range bin while both are moving, this experiment demonstrates simultaneous
moving-target separation rather than the minimum resolvable target spacing.

\subsection{Simultaneous Communication and Sensing}
\label{sec:live_isac}

A commercial UE was attached to the n78 standalone cell with the radar worker
and E2 agent enabled. Downlink performance was obtained from the gNB
medium-access-control statistics. In parallel, the radar worker reported the
occupied sensing resources using the corresponding frame and slot timing.

OAI reports
downlink throughput using its goodput estimate, which serves here as the communication-performance metric.
At MCS~27, the highest MCS sustained in
both configurations, the median downlink goodput was
\SI{172.7}{\mega\bit\per\second} with sensing enabled and
\SI{172.9}{\mega\bit\per\second} with it disabled, a difference of
\SI{-0.1}{\percent}. The corresponding median block error rates (BLERs) were
0.0655 and 0.0648. No measurable difference in these
communication metrics was observed under the matched-MCS~27 condition.

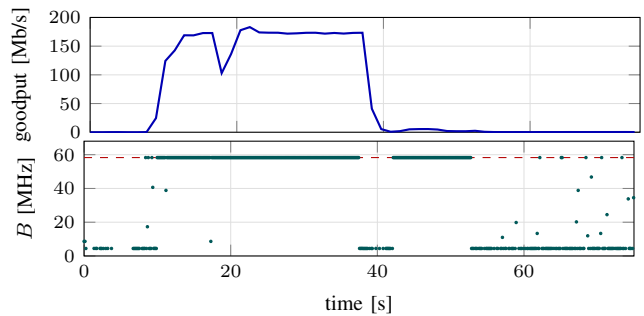
\begin{figure}[ht]
\centering
\begin{tikzpicture}
\begin{axis}[width=\columnwidth, height=3.1cm,
  ylabel={goodput [Mb/s]}, xmin=0, xmax=75, ymin=0, ymax=200,
  xticklabels={}, grid=both, grid style={line width=.2pt, draw=gray!25},
  tick label style={font=\scriptsize}, label style={font=\footnotesize}]
\addplot[color=blue!70!black,thick] coordinates {(0.02,0.00) (1.30,0.00) (2.58,0.18) (3.86,0.22) (5.14,0.03) (6.42,0.01) (7.70,0.29) (8.98,24.59) (10.26,124.28) (11.54,142.62) (12.82,168.94) (14.10,168.80) (15.38,172.79) (16.66,172.86) (17.94,102.98) (19.22,136.23) (20.50,177.76) (21.78,183.18) (23.06,173.94) (24.34,173.45) (25.62,173.39) (26.90,171.80) (28.18,172.35) (29.46,173.27) (30.74,173.36) (32.02,171.82) (33.30,173.00) (34.58,171.94) (35.86,173.09) (37.14,173.34) (38.42,40.98) (39.70,5.46) (40.98,0.84) (42.26,1.90) (43.54,5.08) (44.82,5.49) (46.10,5.55) (47.38,4.72) (48.66,2.11) (49.94,1.77) (51.22,1.73) (52.50,2.52) (53.78,0.77) (55.06,0.34) (56.34,0.27) (57.62,0.26) (58.90,0.29) (60.18,0.24) (61.46,0.32) (62.74,0.32) (64.02,0.30) (65.30,0.34) (66.58,0.32) (67.86,0.32) (69.14,0.34) (70.42,0.33) (71.70,0.35) (72.98,0.38) (74.26,0.32)};
\end{axis}
\end{tikzpicture}\\[-1.5mm]
\begin{tikzpicture}
\begin{axis}[width=\columnwidth, height=3.1cm,
  xlabel={time [s]}, ylabel={$B$ [MHz]}, xmin=0, xmax=75,
  ymin=0, ymax=68, grid=both, grid style={line width=.2pt, draw=gray!25},
  tick label style={font=\scriptsize}, label style={font=\footnotesize}]
\addplot[color=teal!70!black,only marks,mark=*,mark size=0.45pt]
  coordinates {(0.00,8.64) (0.13,8.64) (0.25,4.44) (1.36,4.47) (1.61,4.47) (1.73,4.47) (2.17,4.47) (2.29,4.44) (2.41,4.44) (2.65,4.44) (2.97,4.44) (3.23,4.47) (3.74,4.44) (6.70,4.44) (6.82,4.44) (6.94,4.44) (7.06,4.44) (7.18,4.44) (7.30,4.44) (7.42,4.44) (7.54,4.44) (7.92,4.44) (8.04,4.44) (8.16,4.44) (8.28,4.47) (8.40,58.29) (8.52,4.44) (8.64,17.31) (8.76,58.29) (8.89,4.44) (9.01,4.44) (9.13,4.44) (9.25,58.29) (9.37,40.65) (9.49,4.47) (9.61,4.47) (9.73,4.47) (9.85,4.44) (9.97,58.29) (10.09,58.29) (10.21,58.29) (10.33,58.29) (10.45,58.29) (10.57,58.29) (10.69,58.29) (10.81,58.29) (10.93,58.29) (11.05,58.29) (11.17,38.85) (11.29,58.29) (11.41,58.29) (11.53,58.29) (11.65,58.29) (11.77,58.29) (11.89,58.29) (12.01,58.29) (12.13,58.29) (12.25,58.29) (12.37,58.29) (12.49,58.29) (12.61,58.29) (12.73,58.29) (12.85,58.29) (12.97,58.29) (13.09,58.29) (13.21,58.29) (13.33,58.29) (13.45,58.29) (13.57,58.29) (13.69,58.29) (13.81,58.29) (13.93,58.29) (14.05,58.29) (14.17,58.29) (14.29,58.29) (14.41,58.29) (14.53,58.29) (14.65,58.29) (14.77,58.29) (14.89,58.29) (15.01,58.29) (15.13,58.29) (15.25,58.29) (15.37,58.29) (15.49,58.29) (15.61,58.29) (15.73,58.29) (15.85,58.29) (15.97,58.29) (16.09,58.29) (16.21,58.29) (16.33,58.29) (16.45,58.29) (16.57,58.29) (16.69,58.29) (16.81,58.29) (16.93,58.29) (17.05,58.29) (17.17,58.29) (17.29,8.64) (17.41,58.29) (17.53,58.29) (17.65,58.29) (17.77,58.29) (17.89,58.29) (18.01,58.29) (18.13,58.29) (18.25,58.29) (18.37,58.29) (18.49,58.29) (18.61,58.29) (18.73,58.29) (18.85,58.29) (18.97,58.29) (19.09,58.29) (19.21,58.29) (19.33,58.29) (19.45,58.29) (19.57,58.29) (19.69,58.29) (19.81,58.29) (19.93,58.29) (20.05,58.29) (20.17,58.29) (20.29,58.29) (20.41,58.29) (20.53,58.29) (20.65,58.29) (20.77,58.29) (20.89,58.29) (21.01,58.29) (21.13,58.29) (21.25,58.29) (21.37,58.29) (21.49,58.29) (21.61,58.29) (21.73,58.29) (21.85,58.29) (21.97,58.29) (22.09,58.29) (22.21,58.29) (22.33,58.29) (22.45,58.29) (22.57,58.29) (22.69,58.29) (22.81,58.29) (22.93,58.29) (23.05,58.29) (23.17,58.29) (23.29,58.29) (23.41,58.29) (23.53,58.29) (23.65,58.29) (23.77,58.29) (23.89,58.29) (24.01,58.29) (24.13,58.29) (24.25,58.29) (24.37,58.29) (24.49,58.29) (24.61,58.29) (24.73,58.29) (24.85,58.29) (24.97,58.29) (25.09,58.29) (25.21,58.29) (25.33,58.29) (25.45,58.29) (25.57,58.29) (25.69,58.29) (25.81,58.29) (25.93,58.29) (26.05,58.29) (26.17,58.29) (26.29,58.29) (26.41,58.29) (26.53,58.29) (26.65,58.29) (26.77,58.29) (26.89,58.29) (27.01,58.29) (27.13,58.29) (27.25,58.29) (27.37,58.29) (27.49,58.29) (27.61,58.29) (27.73,58.29) (27.85,58.29) (27.97,58.29) (28.09,58.29) (28.21,58.29) (28.33,58.29) (28.45,58.29) (28.57,58.29) (28.69,58.29) (28.81,58.29) (28.93,58.29) (29.05,58.29) (29.17,58.29) (29.29,58.29) (29.41,58.29) (29.53,58.29) (29.65,58.29) (29.77,58.29) (29.89,58.29) (30.01,58.29) (30.13,58.29) (30.25,58.29) (30.37,58.29) (30.49,58.29) (30.61,58.29) (30.73,58.29) (30.85,58.29) (30.97,58.29) (31.09,58.29) (31.21,58.29) (31.33,58.29) (31.45,58.29) (31.57,58.29) (31.69,58.29) (31.81,58.29) (31.93,58.29) (32.05,58.29) (32.17,58.29) (32.29,58.29) (32.41,58.29) (32.53,58.29) (32.65,58.29) (32.77,58.29) (32.89,58.29) (33.01,58.29) (33.13,58.29) (33.25,58.29) (33.37,58.29) (33.49,58.29) (33.61,58.29) (33.73,58.29) (33.85,58.29) (33.97,58.29) (34.09,58.29) (34.21,58.29) (34.33,58.29) (34.45,58.29) (34.57,58.29) (34.69,58.29) (34.81,58.29) (34.93,58.29) (35.05,58.29) (35.17,58.29) (35.29,58.29) (35.41,58.29) (35.53,58.29) (35.65,58.29) (35.77,58.29) (35.89,58.29) (36.01,58.29) (36.13,58.29) (36.25,58.29) (36.37,58.29) (36.49,58.29) (36.61,58.29) (36.73,58.29) (36.85,58.29) (36.97,58.29) (37.09,58.29) (37.21,58.29) (37.33,58.29) (37.45,58.29) (37.57,4.47) (37.69,4.44) (37.81,4.47) (37.93,4.47) (38.05,4.44) (38.17,4.44) (38.29,4.44) (38.42,4.44) (38.54,4.44) (38.67,4.44) (38.93,4.44) (39.19,4.44) (39.31,4.44) (39.80,4.47) (40.29,4.44) (40.79,4.44) (41.16,4.47) (41.29,4.44) (41.41,4.47) (41.53,4.44) (41.65,4.44) (41.77,4.44) (42.05,4.44) (42.17,58.29) (42.29,58.29) (42.41,58.29) (42.53,58.29) (42.65,58.29) (42.77,58.29) (42.89,58.29) (43.01,58.29) (43.13,58.29) (43.25,58.29) (43.37,58.29) (43.49,58.29) (43.61,58.29) (43.73,58.29) (43.85,58.29) (43.97,58.29) (44.09,58.29) (44.21,58.29) (44.33,58.29) (44.45,58.29) (44.57,58.29) (44.69,58.29) (44.81,58.29) (44.93,58.29) (45.05,58.29) (45.17,58.29) (45.29,58.29) (45.41,58.29) (45.53,58.29) (45.65,58.29) (45.77,58.29) (45.89,58.29) (46.01,58.29) (46.13,58.29) (46.25,58.29) (46.37,58.29) (46.49,58.29) (46.61,58.29) (46.73,58.29) (46.85,58.29) (46.97,58.29) (47.09,58.29) (47.21,58.29) (47.33,58.29) (47.45,58.29) (47.57,58.29) (47.69,58.29) (47.81,58.29) (47.93,58.29) (48.05,58.29) (48.17,58.29) (48.29,58.29) (48.41,58.29) (48.53,58.29) (48.65,58.29) (48.77,58.29) (48.89,58.29) (49.01,58.29) (49.13,58.29) (49.25,58.29) (49.37,58.29) (49.49,58.29) (49.61,58.29) (49.73,58.29) (49.85,58.29) (49.97,58.29) (50.09,58.29) (50.21,58.29) (50.33,58.29) (50.45,58.29) (50.57,58.29) (50.69,58.29) (50.81,58.29) (50.93,58.29) (51.05,58.29) (51.17,58.29) (51.29,58.29) (51.41,58.29) (51.53,58.29) (51.65,58.29) (51.77,58.29) (51.89,58.29) (52.01,58.29) (52.13,58.29) (52.25,58.29) (52.37,58.29) (52.49,58.29) (52.61,58.29) (52.73,58.29) (52.85,4.44) (52.98,4.47) (53.11,4.44) (53.35,4.47) (53.47,4.44) (53.59,4.44) (53.86,4.47) (53.99,4.44) (54.11,4.44) (54.23,4.44) (54.60,4.44) (54.84,4.47) (54.96,4.44) (55.09,4.44) (55.21,4.44) (55.33,4.44) (55.45,4.44) (55.57,4.44) (55.93,4.44) (56.29,4.47) (56.43,4.44) (56.81,4.44) (56.93,4.47) (57.06,11.01) (57.18,4.44) (57.42,4.47) (57.54,4.47) (57.91,4.44) (58.03,4.44) (58.32,4.44) (58.44,4.44) (58.56,4.44) (58.80,4.44) (58.93,19.83) (59.30,4.44) (59.54,4.44) (59.67,4.44) (59.94,4.44) (60.18,4.44) (60.43,4.44) (60.55,4.44) (60.79,4.44) (60.93,4.44) (61.05,4.44) (61.17,4.44) (61.30,4.44) (61.43,4.44) (61.55,4.47) (61.68,4.44) (61.80,13.35) (61.92,4.47) (62.04,4.44) (62.16,58.29) (62.45,4.44) (62.58,4.44) (62.70,4.44) (62.94,4.44) (63.06,4.47) (63.19,4.44) (63.31,4.44) (63.43,4.44) (63.56,4.44) (63.68,4.44) (63.83,4.47) (64.08,4.47) (64.33,4.44) (64.46,4.44) (64.70,4.44) (64.82,4.44) (64.95,4.44) (65.08,58.29) (65.20,58.29) (65.32,4.44) (65.56,4.47) (65.69,4.44) (65.81,4.47) (65.93,4.44) (66.17,4.44) (66.29,4.44) (66.41,4.44) (66.53,4.44) (66.66,4.44) (66.91,4.47) (67.03,4.44) (67.15,20.19) (67.27,4.44) (67.39,38.85) (67.52,4.47) (67.64,4.44) (67.76,4.44) (68.05,4.44) (68.17,4.47) (68.45,58.29) (68.69,11.94) (68.82,4.44) (69.06,4.44) (69.19,46.77) (69.31,4.47) (69.43,4.44) (69.55,4.47) (69.68,4.44) (69.92,4.47) (70.04,4.47) (70.16,4.44) (70.28,4.47) (70.46,13.35) (70.59,58.29) (70.72,4.44) (70.84,4.44) (70.96,4.44) (71.08,4.47) (71.32,24.48) (71.45,4.44) (71.57,4.47) (71.69,4.44) (71.82,4.44) (71.94,4.44) (72.06,4.44) (72.18,4.47) (72.42,4.47) (72.61,4.44) (72.73,4.44) (72.85,4.44) (72.97,4.44) (73.09,4.44) (73.21,4.47) (73.36,58.29) (73.48,4.44) (73.61,4.44) (73.85,4.44) (74.22,33.81) (74.34,4.47) (74.46,4.47) (74.58,4.44) (74.82,4.44) (74.95,34.53) (75.07,4.44) (75.19,4.47) (75.31,4.44) (75.44,4.47)};
\addplot[dashed,red!70!black] coordinates {(0,58.29) (75,58.29)};
\end{axis}
\end{tikzpicture}
\caption{Simultaneous communication and sensing with a scheduled UE.
Upper panel: gNB-reported downlink goodput. Lower panel: occupied sensing
bandwidth for each CPI. Under downlink load, the full sensing bandwidth is
available; during idle periods, it contracts to the control-channel
footprint.} \label{fig:isac} \end{figure}

The principal interaction was instead in the sensing bandwidth.
Figure~\ref{fig:isac} shows that the available radar bandwidth follows the
scheduler allocation. Under downlink load, the scheduler provides the full
162-PRB allocation, retaining \SI{58.29}{\mega\hertz} of sensing bandwidth
and approximately \SI{2.6}{\meter} nominal range resolution. During idle periods, the occupied bandwidth falls to approximately
\SI{4.44}{\mega\hertz}, corresponding to the control-channel footprint and a
nominal range resolution of approximately \SI{33.8}{\meter}. Thus, the
\code{--phy-test} configuration used for the main radar characterization
represents a fully allocated cell but not an idle one.
Table~\ref{tab:isac_alloc} summarizes the corresponding allocation states.

\begin{table}[ht]
\centering
\caption{Sensing Bandwidth Versus Scheduler Allocation}
\label{tab:isac_alloc}
\footnotesize
\setlength{\tabcolsep}{4pt}
\begin{tabular}{L{3.2cm}rrr}
\toprule
Allocation & CPIs & $B$ [MHz] & $\Delta R$ [m]\\
\midrule
Forced full-band allocation & 203 & 58.29 & 2.6\\
Live UE, loaded             & 426 & 58.29 & 2.6\\
Live UE, light              & 358 & 58.29 & 2.6\\
Live UE, idle               & 274 & 4.44  & 33.8\\
\bottomrule
\end{tabular}
\end{table}

\section{Conclusion}
\label{sec:conclusion}

This paper presented a real-time symbol-domain OFDM radar integrated directly
into the OAI 5G gNB and exposed through a custom O-RAN E2 sensing service. The
same processing chain was validated progressively in an offline physical-layer
simulator, the OAI RF simulator, and a USRP X300 hardware implementation,
providing a continuous path from controlled verification to live over-the-air
operation.

Hardware measurements identified a deterministic carrier-dependent TX--RX
phase rotation on the X300 that can severely limit coherent processing if left
uncorrected. At the selected operating carrier, the measured residual rotation
was reduced from \SI{28.5714}{\hertz} to \SI{0.0012}{\hertz}, improving
mean-removal clutter suppression from \SI{-16.4}{\decibel} to
\SI{38.0}{\decibel} and reducing the $M=64$ coherent-integration loss from
\SI{19.81}{\decibel} to \SI{0.27}{\decibel}. At this operating point, the
measured processing gain of \SI{57.62}{\decibel} closely matched the predicted
\SI{57.56}{\decibel}. The resulting range--Doppler processing supported
moving-target detection, range and Doppler estimation, and simultaneous
separation of two oppositely moving targets.

Instrumented execution measurements confirmed that the radar processing fits
within the available real-time budget. The combined per-sounding and per-CPI
processing occupied \SI{45.2}{\percent} of the CPI interval using median stage
times, while a conservative bound constructed from the separately observed
stage maxima was \SI{58.4}{\percent}. The maximum input-ring occupancy was 22
of 64 soundings, and no soundings were dropped during the instrumented run.

The E2 implementation further demonstrated that the sensing products can be
exported from the gNB as network-visible measurements. During a
\SI{416}{\second} live session, 2591 CPI reports were delivered without a
missing CPI sequence step or radar-ring drop. The exported detections and
compact slow-time payload were sufficient for controller-side tracking,
micro-Doppler processing, recording, and classification. Operation with a
commercial UE on the same carrier also showed no measurable communication
degradation under the tested matched-MCS~27 condition: median downlink goodput
was \SI{172.7}{\mega\bit\per\second} with sensing enabled and
\SI{172.9}{\mega\bit\per\second} with sensing disabled. In contrast, the available sensing bandwidth remained coupled to
the communication scheduler: a loaded cell provided the full
\SI{58.29}{\mega\hertz} occupied bandwidth, whereas reduced allocation during
idle periods directly degraded range resolution.

Overall, the results demonstrate that established symbol-domain OFDM radar can
operate as a real-time sensing function inside a 5G base station, reuse the
scheduled communication waveform without introducing a separate radar
transmission, and expose sensing products through an O-RAN control path. They
also highlight that practical cellular sensing is a system-level problem:
radar performance depends not only on the signal-processing algorithm, but
also on radio coherence, bounded real-time execution and buffering, and the
communication scheduler that determines the instantaneous sensing resources.

\section{Acknowledgment}
The authors acknowledge the use of Anthropic models as coding and writing assistants and OpenAI models as writing assistants during the preparation of this manuscript. The authors are committed to the responsible and transparent use of AI-assisted tools and take full responsibility for the accuracy, integrity, and content of the paper.

\bibliographystyle{IEEEtran}
\bibliography{references}

@INPROCEEDINGS{IC_idea_first,
  author={Sit, Yoke Leen and Sturm, Christian and Zwick, Thomas},
  booktitle={2011 8th European Radar Conference}, 
  title={Interference cancellation for dynamic range Improvement in an OFDM joint radar and communication system}, 
  year={2011},
  volume={},
  number={},
  pages={333-336},
  doi={}}

@article{OAI_foundational,
title = {Driving innovation in 6G wireless technologies: The OpenAirInterface approach},
journal = {Computer Networks},
volume = {269},
pages = {111410},
year = {2025},
issn = {1389-1286},
doi = {10.1016/j.comnet.2025.111410},
url = {https://www.sciencedirect.com/science/article/pii/S1389128625003779},
author = {Florian Kaltenberger and Tommaso Melodia and Irfan Ghauri and Michele Polese and Raymond Knopp and Tien Thinh Nguyen and Sakthivel Velumani and Davide Villa and Leonardo Bonati and Robert Schmidt and Sagar Arora and Mikel Irazabal and Navid Nikaein}
}

@ARTICLE{PRS_fundamentals,
  author={Wei, Zhiqing and Wang, Yuan and Ma, Liang and Yang, Shaoshi and Feng, Zhiyong and Pan, Chengkang and Zhang, Qixun and Wang, Yajuan and Wu, Huici and Zhang, Ping},
  journal={IEEE Transactions on Vehicular Technology}, 
  title={5G PRS-Based Sensing: A Sensing Reference Signal Approach for Joint Sensing and Communication System}, 
  year={2023},
  volume={72},
  number={3},
  pages={3250-3263},
  doi={10.1109/TVT.2022.3215159}}

@INPROCEEDINGS{SW1,
  author={Sturm, Christian and Zwick, Thomas and Wiesbeck, Werner},
  booktitle={VTC Spring 2009 - IEEE 69th Vehicular Technology Conference}, 
  title={An OFDM System Concept for Joint Radar and Communications Operations}, 
  year={2009},
  volume={},
  number={},
  pages={1-5},
  doi={10.1109/VETECS.2009.5073387}}

@INPROCEEDINGS{SW_MLE,
  author={Braun, Martin and Sturm, Christian and Jondral, Friedrich K.},
  booktitle={2010 IEEE Radar Conference}, 
  title={Maximum likelihood speed and distance estimation for OFDM radar}, 
  year={2010},
  volume={},
  number={},
  pages={256-261},
  doi={10.1109/RADAR.2010.5494616}}

@ARTICLE{SW_invited_paper,
  author={Sturm, Christian and Wiesbeck, Werner},
  journal={Proceedings of the IEEE}, 
  title={Waveform Design and Signal Processing Aspects for Fusion of Wireless Communications and Radar Sensing}, 
  year={2011},
  volume={99},
  number={7},
  pages={1236-1259},
  doi={10.1109/JPROC.2011.2131110}}

@INPROCEEDINGS{SW_performance_verification,
  author={Sturm, Christian and Zwick, Thomas and Wiesbeck, Werner and Braun, Martin},
  booktitle={2010 IEEE Radar Conference}, 
  title={Performance verification of symbol-based OFDM radar processing}, 
  year={2010},
  volume={},
  number={},
  pages={60-63},
  doi={10.1109/RADAR.2010.5494651}}

@misc{adi_grofdmradar,
  author       = {{Analog Devices Inc.}},
  title        = {gr-ofdmradar: {OFDM} Radar on {MxFE} Platforms using {IIO}},
  howpublished = {Analog Devices Wiki,
                  \url{https://wiki.analog.com/resources/eval/user-guides/ad9081_fmca_ebz/radar}},
  year         = {2023},
  note         = {Based on a GRCon 2021 presentation; code:
                  \url{https://github.com/analogdevicesinc/gr-ofdmradar}.
                  Last modified 30 Jan 2023, accessed July 2026}
}

@misc{batstation,
      title={{BatStation}: Toward In-Situ Radar Sensing on {5G} Base Stations with Zero-Shot Template Generation}, 
      author={Zhihui Gao and Zhecun Liu and Tingjun Chen},
      year={2025},
      eprint={2509.06898},
      archivePrefix={arXiv},
      primaryClass={cs.NI},
      url={https://arxiv.org/abs/2509.06898}, 
}

@misc{cellsense,
      title={CellSense: A Sub-6 GHz Cellular ISAC System for Clutter-Robust Passive Sensing}, 
      author={Bibhor Kumar and Ish Kumar Jain and Vijay K Shah},
      year={2026},
      eprint={2606.07900},
      archivePrefix={arXiv},
      primaryClass={eess.SY},
      url={https://arxiv.org/abs/2606.07900}, 
}

@article{dapp_framework,
title = {dApps: Enabling real-time AI-based Open RAN control},
journal = {Computer Networks},
volume = {269},
pages = {111342},
year = {2025},
issn = {1389-1286},
doi = {10.1016/j.comnet.2025.111342},
url = {https://www.sciencedirect.com/science/article/pii/S1389128625003093},
author = {Andrea Lacava and Leonardo Bonati and Niloofar Mohamadi and Rajeev Gangula and Florian Kaltenberger and Pedram Johari and Salvatore D’Oro and Francesca Cuomo and Michele Polese and Tommaso Melodia}
}

@misc{dapp_isac,
      title={Enabling Programmable Inference and ISAC at the 6GR Edge with dApps}, 
      author={Michele Polese and Rajeev Gangula and Tommaso Melodia},
      year={2026},
      eprint={2603.29146},
      archivePrefix={arXiv},
      primaryClass={cs.NI},
      url={https://arxiv.org/abs/2603.29146}, 
}

@article{fd_ofdm_radar_tmtt,
   title={Full-Duplex OFDM Radar With LTE and 5G NR Waveforms: Challenges, Solutions, and Measurements},
   volume={67},
   ISSN={1557-9670},
   url={http://dx.doi.org/10.1109/TMTT.2019.2930510},
   DOI={10.1109/tmtt.2019.2930510},
   number={10},
   journal={IEEE Transactions on Microwave Theory and Techniques},
   publisher={Institute of Electrical and Electronics Engineers (IEEE)},
   author={Baquero Barneto, Carlos and Riihonen, Taneli and Turunen, Matias and Anttila, Lauri and Fleischer, Marko and Stadius, Kari and Ryynanen, Jussi and Valkama, Mikko},
   year={2019},
   month=Oct, pages={4042–4054} }

@misc{gnuradio,
    author       = {{GNU Radio project}},
    title        = {{GNU Radio}: The Free \& Open Source Radio Ecosystem},
    howpublished = {\url{https://www.gnuradio.org}},
    note         = {Version 3.10, accessed July 2026}
  }

@INPROCEEDINGS{mimo_jrc_mmwave,
  author={Ozkaptan, Ceyhun D. and Zhu, Haocheng and Ekici, Eylem and Altintas, Onur},
  booktitle={2023 IEEE 3rd International Symposium on Joint Communications \& Sensing (JC\&S)}, 
  title={Software-Defined MIMO OFDM Joint Radar-Communication Platform with Fully Digital mmWave Architecture}, 
  year={2023},
  volume={},
  number={},
  pages={1-6},
  doi={10.1109/JCS57290.2023.10107511}}

@INPROCEEDINGS{oai_isac_ew25,
  author={Carbonara, Salvatore and Pugliese, Daniele and Fascista, Alessio and Coluccia, Angelo and Boggia, Gennaro},
  booktitle={European WIRELESS 2025; 30th European Wireless Conference}, 
  title={5G-Compliant Integrated Sensing and Communication at Sub-6 GHz: Experiments with SDRs and OpenAirInterface}, 
  year={2025},
  volume={},
  number={},
  pages={232-237},
  doi={}}

@INPROCEEDINGS{oai_isac_jcs26,
  author={Carbonara, Salvatore and Pugliese, Daniele and Fascista, Alessio and Coluccia, Angelo and Boggia, Gennaro},
  booktitle={2026 IEEE 6th International Symposium on Joint Communications \& Sensing (JC\&S)}, 
  title={Downlink ISAC with a Full-Stack 5G-Compliant Experimental Testbed: Communication vs. Control Signals for Multi-Target Detection}, 
  year={2026},
  volume={},
  number={},
  pages={1-6},
  doi={10.1109/JCS69321.2026.11366013}}

@INPROCEEDINGS{ofdm_mimo_angular,
  author={Sit, Yoke Leen and Nguyen, Thuy T. and Sturm, Christian and Zwick, Thomas},
  booktitle={2013 European Radar Conference}, 
  title={2D radar imaging with velocity estimation using a MIMO OFDM-based radar for automotive applications}, 
  year={2013},
  volume={},
  number={},
  pages={145-148},
  doi={}}

@INPROCEEDINGS{ofdm_multiuser,
  author={Sit, Yoke Leen and Reichardt, Lars and Sturm, Christian and Zwick, Thomas},
  booktitle={2011 IEEE RadarCon (RADAR)}, 
  title={Extension of the OFDM joint radar-communication system for a multipath, multiuser scenario}, 
  year={2011},
  volume={},
  number={},
  pages={718-723},
  doi={10.1109/RADAR.2011.5960632}}

@INPROCEEDINGS{ofdm_sturm_first,
  author={Sturm, C. and Pancera, E. and Zwick, T. and Wiesbeck, W.},
  booktitle={2009 IEEE Radar Conference}, 
  title={A novel approach to OFDM radar processing}, 
  year={2009},
  volume={},
  number={},
  pages={1-4},
  doi={10.1109/RADAR.2009.4977002}}

@misc{openisac,
      title={OpenISAC: An Open-Source Real-Time Experimentation Platform for OFDM-ISAC}, 
      author={Zhiwen Zhou and Chaoyue Zhang and Xiaoli Xu and Yong Zeng},
      year={2026},
      eprint={2601.03535},
      archivePrefix={arXiv},
      primaryClass={eess.SP},
      url={https://arxiv.org/abs/2601.03535}, 
}

@ARTICLE{oran_foundational,
  author={Polese, Michele and Bonati, Leonardo and D’Oro, Salvatore and Basagni, Stefano and Melodia, Tommaso},
  journal={IEEE Communications Surveys \& Tutorials}, 
  title={Understanding {O-RAN}: Architecture, Interfaces, Algorithms, Security, and Research Challenges}, 
  year={2023},
  volume={25},
  number={2},
  pages={1376-1411},
  doi={10.1109/COMST.2023.3239220}}

@misc{oran_isac_arch_analysis2,
      title={Toward Native ISAC Support in O-RAN Architectures for 6G}, 
      author={Eduardo Baena and Rajesh Krishnan and Mai Vu and Gil Zussman and Dimitrios Koutsonikolas},
      year={2026},
      eprint={2603.03607},
      archivePrefix={arXiv},
      primaryClass={cs.NI},
      url={https://arxiv.org/abs/2603.03607}, 
}

@misc{oran_loc_e2sm_srs,
      title={An {O-RAN} Framework for {AI/ML}-Based Localization with {OpenAirInterface} and {FlexRIC}}, 
      author={Nada Bouknana and Mohsen Ahadi and Florian Kaltenberger and Robert Schmidt},
      year={2026},
      eprint={2511.19233},
      archivePrefix={arXiv},
      primaryClass={cs.NI},
      url={https://arxiv.org/abs/2511.19233}, 
}

@inproceedings{schmidt2021flexric,
author = {Schmidt, Robert and Irazabal, Mikel and Nikaein, Navid},
title = {FlexRIC: an SDK for next-generation SD-RANs},
year = {2021},
isbn = {9781450390989},
publisher = {Association for Computing Machinery},
address = {New York, NY, USA},
url = {https://doi.org/10.1145/3485983.3494870},
doi = {10.1145/3485983.3494870},
booktitle = {Proceedings of the 17th International Conference on Emerging Networking EXperiments and Technologies},
pages = {411–425},
numpages = {15},
location = {Virtual Event, Germany},
series = {CoNEXT '21}
}

@ARTICLE{senseoran,
  author={Reus-Muns, Guillem and Upadhyaya, Pratheek S. and Demir, Utku and Stephenson, Nathan and Soltani, Nasim and Shah, Vijay K. and Chowdhury, Kaushik R.},
  journal={IEEE Journal on Selected Areas in Communications}, 
  title={{SenseORAN}: {O-RAN} Based Radar Detection in the {CBRS} Band}, 
  year={2024},
  volume={42},
  number={2},
  pages={326-338},
  doi={10.1109/JSAC.2023.3336152}}

@online{uhd_tuning_notes,
  author  = {{Ettus Research / National Instruments}},
  title   = {{USRP} Hardware Driver and {USRP} Manual: General Application
             Notes --- Tuning Notes},
  year    = {2024},
  url     = {https://uhd.readthedocs.io/en/latest/page_general.html},
  urldate = {2026-08-05}
}

\end{document}